\documentclass[12pt,english,preprint]{aastex}

\usepackage[T1]{fontenc}
\usepackage[latin9]{inputenc}
\usepackage{amstext}
\usepackage{amsmath}
\usepackage{amssymb}

\usepackage{graphicx}
\usepackage{color}
\usepackage{natbib}

\usepackage{xfrac}
\usepackage{diagbox}
\usepackage{multirow}

\usepackage{longtable}

\usepackage{babel}

\usepackage{xifthen}

\doublespace

\newcommand{\txt}{\text}
\newcommand{\mybf}{}

\newcommand{\su}[2]{#1_{\rm #2}}

\newcommand{\mdt}[1][]{ 
  \ifthenelse{\isempty{#1}}
  {\dot{M}_{\rm loc}}
  { {\dot M}_{\rm{loc},{#1}} } 
  } 
  
\newcommand{\Mdt}[1][]{ 
    \ifthenelse{\isempty{#1}}
    {\dot{M}_{}}
    { {\dot M}_{{#1}} } 
} 

\newcommand{\Rout}{\su{R}{out}}
\newcommand{\Rin}{\su{R}{in}}

\newcommand{\mdtcr}{\su{\dot{M}}{loc,cr} }

\newcommand{\MsolYrM}{ \,M_{\odot}{\rm yr}^{-1} }

\newcommand{\rsub}{ \su{R}{sub} }

\newcommand{\Tsub}{ \su{T}{sub} }
\newcommand{\Tc}{T}
\newcommand{\Tcgas}{\su{T}{c,gas}}

\usepackage{babel}

\usepackage{babel}
\begin{document}
\global\long\def\pd{\partial}%

\author{A. Dorodnitsyn\altaffilmark{1,2,3}, T. Kallman\altaffilmark{1}}
\altaffiltext{1}{Laboratory for High Energy Astrophysics, NASA Goddard Space 
Flight Center, Code 662, Greenbelt, 
MD, 20771, USA}
\altaffiltext{2}{University of Maryland, Baltimore County (UMBC/CRESST), Baltimore, MD 21250, USA}
\altaffiltext{3}{Space Research Institute, 84/32, Profsoyuznaya st., Moscow, Russia}

\title{A physical model for radiative, convective dusty disk in AGN.}

\begin{abstract}
  
An accretion disk in an Active Galactic Nucleus (AGN) harbors and shields dust from external
illumination: at the mid-plane of the disk around a $M_{{\rm
BH}}=10^{7}M_{\odot}$ black hole, dust can exist 
at $0.1$pc from the black hole, compared to 0.5pc outside of the disk.  
We construct a physical model of a disk region 
approximately located  between the radius of dust sublimation at the
disk mid-plane and the radius at which dust sublimes at the disk
surface. Our main conclusion is that for a wide range of
model parameters such as local accretion rate and/or opacity, the accretion 
disk's own radiation pressure on dust significantly influences its vertical structure.  In addition to being highly convective, such a disk can transform from geometrically thin to slim.  Our model fits into the narrative of a "failed wind" scenario of
\citet{CzernyHryniewicz11} and the "compact torus" model of
\citet{BaskinLaor2018MNRAS}, incorporating them as variations of the radiative dusty disk model.
\end{abstract}

\section{Introduction}\label{sec:Introduction}

The radiative output of Active Galactic Nuclei (AGN) is powered
by accretion of gas and most of the gas potential energy is
released in the inner part of an accretion disk.  However, several crucial observational
characteristics of AGN are shaped considerably further away, at few$\times0.01-0.5$pc
from the Super-Massive Black Hole (BH).  It is also at approximately  this
distance from the BH the radiation flux is sufficiently diluted such
that dust grains can survive the illumination from the nucleus.
The presence of dust results in a $10-100$ -fold increase of opacity 
compared with only gas, which leads to a  dramatic increase of coupling between the radiation
from the nucleus and gas.  

The radius outside of which dust can survive forms a dust sublimation
surface, a boundary between the inner, mostly
dust-free region and the outer part often associated with the dusty
torus.  The latter is
invoked to to explain the dichotomy between two types of AGN, in which 
optically thick equatorial material blocks the direct view onto the broad line region
and the accretion disk in type 2 galaxies
\citep{Rowan-Robinson77,Antonucci84,AntonucciMiller1985,UrryPadovani95}.

Mid-infrared (MIR) interferometric observations of nearby AGN 
clearly point to the presence of dust
\citep{Jaffe2004,Raban09,Tristram14,Tristram2011} 
at distances $\geq$ 0.1 pc from the center.  The relative numbers of type 
1 and type 2 objects suggest that the obscuring material is geometrically thick.

The virial theorem predicts
that in order to be geometrically thick at a distance, $r\simeq1\,\text{pc}$,  
temperature of the obscuring gas should be of the order of $10^{6}\text{K}$ for
a $10^{7}M_{\odot}$ BH. This is not compatible with survival of dust in the obscurer,
and hence is in conflict with the presence of dust inferred from IR observations.
On the other hand, the temperature expected at the surface of a thin accretion disk 
is $\leq$1000 K at $\sim$1 pc from the center.

Reverberation mapping \citep[i.e.][]{Koshida2014} is generally consistent
 with putting the inner boundary of the torus to 
within the dust sublimation surface at $0.4-0.5$pc \citep{Kaspi2000}.
The location of the broad line region (BLR) relative to the center has been measured 
by reverberation to be $\sim \text{few}\times\,0.01$ pc 
\citep[e.g.][]{Peterson2004,Suganuma06}.

Magnetic or/and radiation driving have been proposed as a mechanisms behind the formation of the BLR and the torus.
A line driven wind, i.e. a wind driven by the radiation pressure in UV lines \citep{ProgaKallman2004,Murr05}, 
has a launching radius which is $\sim 10x$ smaller, and correspondingly characteristic 
line widths which are a factor $\sim 10x$ greater than indicated by the maximum BLR line widths. A line driven wind from an accretion disk is not massive enough to be the torus. 

Magnetic fields, specifically large-scale magnetic fields, 
are an alternative or augmenting mechanism which can be the driving engine of the BLR and the torus. 
Semi-analytical or numerical models \citep{Lovelace98,Dorodnitsyn16} 
show that large-scale B-field can support an AGN torus. 
Self-consistent numerical simulations of a 
thin disk threaded by the net vertical magnetic flux \citep{ZhuStone2018} 
show that thin disks cannot both transport large-scale magnetic fields 
\citep{LubowPapaloizou94,BKLovelace2000,BKLovelace2007} \emph{and} simultaneously have a 
massive, polar, MHD-driven (Magneto-hydrodynamic) outflow. 
In addition,  typical MHD flows have an approximate equipartition between magnetic energy density and 
gas energy. So the characteristic temperature of the gas near the launching radius of the magnetically-driven, 
gravitationally unbound outflow is also expected be of the order of the virial temperature.

\citet{CzernyHryniewicz11}(hereafter CH11) suggested that the disk's own radiation may produce sufficient
radiation pressure on dust grains that would expel a failed wind from the outer accretion disk atmosphere. 
Such a failed wind then would be responsible for the formation of the broad line region seen
in Seyfert I galaxies. Developing this idea further, \citet{BaskinLaor2018MNRAS}
(BL18) concluded that the contribution from large graphite grains
near $T\simeq2000{\rm K}$ increases dust opacity which results in
a an inflated compact, torus-like structure near the observed BLR radius. 

In  a simple case, when dust is arranged in a spherically-symmetric
shell it cannot survive closer than dust sublimation radius: $R_{\text{sub}}\simeq0.2-0.5$
pc from a supermassive black hole.
If dust is contained in a cold and dense accretion disk it can survive
much closer to the BH, down to $\sim10^{-3}-10^{-2}$ pc (BL18, CH11). The radial
scaling of the effective surface temperature of the disk: $T_{{\rm eff}}\propto R^{-3/4}$
guarantees that beyond a certain radius, $R_{{\rm din}}$, $T$ drops
below the dust sublimation temperature
allowing survival of dust. This corresponds to 
to a dramatic increase of the opacity. The disk
radiation flux
may be strong enough to produce a non-negligible radiation pressure
$\propto(F/c)\,\kappa_{d}$, where $\kappa_{d}$ is dust opacity.
As  suggested by CH11 and  BL18 this can lead to the formation of failed
winds or produce a "compact torus" at $\text{few}\times10^{-2}$pc.


Previous work has demonstrated the likely importance of dust in the dynamics of the gas in the 
torus and BLR.  However, there has not been an examination of the effects of dust on the internal structure of 
the accretion disk in the pc-scale region of AGN. 
The goal of this paper is to explore how radiation pressure on dust grains defines 
the vertical structure of the disk in the region where its temperature is comparable to dust 
sublimation temperature, $\Tsub$.
We will show that such changes are important, and lead to verifiable
results. A preview of the results is as follows:

\begin{itemize}
    \item  We first will develop an analytical model of an accretion disk, based on a modified 
    $\alpha$ disk approach, that includes  contributions from gas pressure 
    and radiation pressure on dust.   We show that there is a difference between the  
    radius of dust sublimation at the disk mid-plane and the radius at which dust 
    sublimes at the disk surface.  

    \item There is a region in the AGN accretion disk where the radiation pressure from the disk's own near- and mid-infrared 
    radiation shapes the vertical disk structure. In this paper we call this part of the disk the "Active Dusty Region" (ADR).

    \item It is well known that radiation pressure often leads to a strong convective instability both in stellar envelopes 
    and in radiative disks.   Developing a semi-analytical disk model,
    we calculate the parameter range where convection develops due to vertical radiation pressure on dust and calculate 
    the disk properties. 

    \item The internal radiation push 
    can provide an explanation of the geometrically
    thick obscuration and avoids the  apparent paradoxes associated with 
    gas pressure or turbulent vertical 
    support.
    The main parameter which determines the importance of local radiation pressure is the local accretion rate $\mdt$. 
    The disk becomes locally geometrically thick if the accretion is locally super-Eddington, with the latter calculated 
    with dust opacity which is 10-100 times larger than the typical opacity of gas without dust. 
    The latter allows to make an argument that the obscuration associated with type-2 AGN can be 
    attributed to a supper-critical dusty accretion in the ADR region, calculated in this paper.  
\end{itemize}

The structure of this paper is as follows. In Section \ref{sec:Properties} we  
begin to examine the effects of the disk's own radiation pressure on
dust on the disk vertical structure.
In Section \ref{sec:DustSubRegIndisk} we further derive properties of  the Active Disk Region.
In Section \ref{sec:Solution} we calculate a detailed analytical and numerical solution of 
the disk equations and in 
Section \ref{sec:Convection} we show that in a wide range of parameters
the radiative dusty disk is convectively unstable. The results are summarized in Section \ref{sec:Discussion}
along with the limitations of our approach, and
the ideas developed in this paper are put in the context of the broader AGN accretion disk physics.
We conclude with Section \ref{sec:Conclusions}.

\section{Properties of AGN at pc-scales}\label{sec:Properties}

In this section we examine the effect of dust formation or survival and its dependence on the global parameters of the 
AGN.   
In doing so, we adopt standard assumptions about accretion in a thin disk around a BH
\citep{ShakuraSunyaev73,Lynden-BellPringle74}.

\subsection{Global parameters }\label{sec:GlobalParameters}

It is customary to express the total luminosity of the AGN in terms of the accretion rate:

\begin{equation}
L=\epsilon\:\Mdt c^{2},\label{eq:AGNLum1}
\end{equation}

\noindent which is equivalent to the definition of an accretion efficiency, $\epsilon$ -
a parameter that is approximately bounded between $0.057$
for a non-rotating BH and $0.42$ for a BH rotating at maximum efficiency.
The mass-accretion rate, ${\dot{M}}$ in (\ref{eq:AGNLum1}) corresponds
to the innermost part of the disk where the most of the radiative output
is produced. After assuming $\epsilon$, it is standard to equate (\ref{eq:AGNLum1}) to the Eddington luminosity:

\begin{equation}
L_{{\rm E}}=\frac{4\pi cGM}{\kappa_{e}}=1.25\times10^{45}
M_{7}\text{\text{\,erg} }\,{\rm s}^{-1},\label{eq:EddLum1}
\end{equation}

\noindent where $M$ is the mass of the BH, and to scale accretion rate in terms
of ``Eddington'' accretion rate:

\begin{equation}
\dot{M}_{{\rm E}}\simeq0.22\epsilon_{0.1}^{-1}M_{7}M_{\odot}{\rm yr}^{-1}\mbox{,}\label{eq:EddMdot}
\end{equation}

\noindent where in (\ref{eq:EddLum1}) and (\ref{eq:EddMdot}) the following
parameters are adopted: $\kappa_{e}=0.4\,{\rm cm^{2}g^{-1}}$ is the
Thomson opacity; we scale the BH mass in units of $M_{7}=M/10^{7}\text{M}_{\odot}$
and fix the efficiency of accretion $\epsilon=0.1$. 

The above picture is augmented with the  assumption that enough medium
is supplied to the AGN from galactic scales near the AGN outer
radius, $R_{{\rm AGN}}$. It is customary to define this as a radius
where the gravity from the BH dominates the gravitational field
of the host galaxy:

\begin{equation}
R_{{\rm AGN}}=\frac{GM}{\sigma_{{\rm Blg}}^{2}}\simeq4.30\,M_{7}\left(\frac{\sigma_{{\rm Blg}}{\rm (km}{\rm s^{-1}})}{100}\right)^{-2}\text{ pc}=4.5\times10^{6}
\left(\frac{\sigma_{{\rm Blg}}{\rm (km}{\rm s^{-1}})}{100}\right)^{-2}\,{R_{g}},\label{eq:rAGN}
\end{equation}

\noindent where $\sigma_{{\rm Blg}}$ is the stellar velocity dispersion in
the bulge, and the last equality is given in terms of the Schwarzschild gravitational
radius of the BH: 

\begin{equation}
R_{g}=\frac{2G M}{c^{2}}=2.95\times10^{12}M_{7}\,{\rm cm.}
\label{eq:RadSchwarz}
\end{equation}

\noindent Hereafter, we reserve $R$ for the radius in physical units, and 
$r$ for the scaled radius: $r_y=R/y $.

A crude estimate for the vertical scale height of the disk at $R_{{\rm AGN}}$
is done assuming the scaling for the disk height: $H\sim v_{{\rm T}}/\Omega,$
where $H$ is the half-thickness of the disk, $\Omega=(G M R^{-3})^{1/2}$ is the orbital
velocity and $v_{{\rm T}}$ is the isothermal sound speed, $\su{v}{T}=(\su{{\cal R}}{gas} T/\mu_m)^{1/2}$,
and $\mu_m$ is the mean molecular weight and $\su{{\cal R}}{gas}$ is the gas
constant. Hereafter while calculating disk properties
we neglect the disk self-gravity, and 
from equation (\ref{eq:H2R_at_rAGN}) it follows that at $\su{R}{AGN}$ 
the disk is very thin:

\begin{equation}
H/R_{\text{AGN}} \simeq6.45\times10^{-3}\sigma_{{\rm Blg,}100}^{-1}T_{50}^{1/2}.
\label{eq:H2R_at_rAGN}
\end{equation}
 
 \noindent Consequently, such a disk intercepts only a small fraction
of the radiation flux from the nucleus. 

It is instructive to review the
radiative energy density generated locally in the disk and compare
it to that from external illumination. 
Radiation flux at pc-scales is dominated by the flux from the nucleus, $\su{F}{ext}$
which is produced in the 
inner parts of the disk. Its
angular dependence is the manifestation of the limb darkening effect in the disk, and has a simple
dependence on $\mu=\cos\theta$ where $\theta$ is 
the inclination angle from the normal to the disk
\citep[i.e.][]{sobolevCourseTheoreticalAstrophysics1975, 
sunyaevComptonizationLowfrequencyRadiation1985}:

\begin{equation}
F_{{\rm ext}}\simeq6\times10^{9}f(\theta){\dot{M}}_{0.1}\epsilon_{0.1}\ 
r_{0.1}{}^{-2}\,\,{\rm erg}{\rm \,cm^{-2}{\rm \,s^{-1}}}\gg F_{{\rm loc}}\label{eq:FluxAGNanisotr}
\end{equation}
where $f(\theta)=\mu(2\mu+1)$ is the angular dependence of the radiation flux.
The local radiation flux generated in the disk, i.e. the normal flux at the disk's photosphere reads

\begin{equation}
F_{{\rm loc}}=\frac{3}{8\pi}\frac{GM}{r^{3}}\mdt\simeq3.4\times10^{4}
\,r_{0.1}^{-3}\mdt[0.1] M_{7}{\rm \,erg}\cdot{\rm cm^{-2}}\cdot{\rm s}^{-1}\mbox{,}\label{eq:DiskLocFlux}
\end{equation}

\noindent and it follows that until matter can spiral down to a fraction of a parsec, the release of the
gravitational potential energy produces local radiative output that
is negligible for the gas dynamics. Radiation flux, $\su{F}{ext}$ depends on the mass-accretion rate in the inner disk, 
while $F_{{\rm loc}}$ depends on local accretion rate, and in the following we reserve $\Mdt$ for global, and $\mdt$ for local 
accretion rates.

If a thin, cold, dusty disk is illuminated by the UV flux $F_\txt{ext}$, 
due to very high UV opacity the radiation is stopped immediately near the surface of the disk
heating dust to approximately $\Tsub$. Assuming all incoming radiation is converted to IR it follows 
that the temperature in such an idealized cold disk is {\it decreasing} towards the equatorial plane.
The vertical component of radiation pressure, ${\bf g}_{z,\txt{rad}}\propto (\nabla T)_z$ then 
points downwards and 
is balanced by the vertical gradient of the gas pressure at the characteristic density: 

\begin{equation}
\su{n}{eq} =n(P_g=P_r) \simeq 6.12 \times 10^{10}\,T_{1500}^{3}\,{\rm cm^{-3}}\mbox{,}
\label{eq:densPgas=Prad}    
\end{equation}

\noindent where $T_{1500}$ is the dust temperature scaled in units of $\Tsub=1500$K, also $P_g$ is the gas pressure:

\begin{equation}
P_{g} = \rho{\cal R}T\mbox{,}\label{eq:Pgas}\\
\end{equation}
and $P_r$ is the radiation pressure:  
\begin{equation}
P_r  =  aT^{4}/3\mbox{,}\label{eq:Prad}
\end{equation}

\noindent where $a$ is the radiation constant, and to simplify notation such as in (\ref{eq:Pgas}), the mean molecular weight,
$\mu_m$ is absorbed in the definition of the gas constant,
${\cal R}=\su{{\cal R}}{gas}/\mu_m$, and in the rest of the
paper we adopt $\mu_m=1$ (see also Glossary).  Even when radiation pressure on dust is important in the bulk of the disk, at mid-plane the density exceeds the density in equation \ref{eq:densPgas=Prad}, 
$n_c\gg \su{n}{eq}$,  so despite the enhanced opacity due to dust the mid-plane pressure is dominated by $P_{g}$.

Accretion in a thin disk far  from
a BH is slow. The free-fall time-scale is the shortest in the hierarchy of time-scales: 
$\displaystyle t_{{\rm dyn}}=1.49\times10^{2}r_{0.1}^{3/2}{\rm yr}$. 
However, the accretion time-scale corresponds to the viscous time-scale $t_{{\rm visc}}$ in the disk:

\begin{equation}
  t_{{\rm a}}\propto t_{{\rm visc}}=\frac{R}{v_{r}}=
  5.15\times10^{7}r_{0.1}^{1/2}T_{1500}^{-1}\alpha_{0.1}^{-1}\,\text{yr},\label{eq:t_visc}
\end{equation}

\noindent where $\alpha<1$ is the effective viscosity parameter, introduced
by \cite{ShakuraDiskModelGasAccretionRelativistic1972},
and $\su{\alpha}{0.1} = \alpha/0.1$. 
A geometrically thin disk cools efficiently through
radiative losses and as the disk cooling-time is much smaller
than $t_{a}$: 
$\displaystyle t_{{\rm th}}=1/(\alpha\,\Omega)=1.492\times10^{3}\alpha_{0.1}^{-1}r_{0.1}^{3/2}{\rm yr}$,
it is  approximately $\su{t}{dyn}\lesssim \su{t}{th}$.

From \eqref{eq:t_visc} it is clear that a buildup of matter (in a thin disk) at large radii 
cannot quickly propagate through the disk towards smaller radii. 
Hence we allow the situation in which the local accretion rate exceeds the 
rate at the center, $\mdt\ne\Mdt$. 
The role of the local accretion rate $\mdt$ is to define the local 
rate of energy production in the disk, and 
correspondingly the local vertical radiation flux, $\su{F}{loc}$ in \eqref{eq:DiskLocFlux}.


\subsection{Disk thickness When Radiation Pressure is Important}\label{sec:DiskThicknessWithRadPressure}

Assuming a thin disk
approximation the equation for vertical balance reads:

\begin{equation}
\frac{dP_g}{dz}=-\rho\,g_{z} + \rho\frac{\su{F}{loc}\kappa}{c}\mbox{,}\label{eq:dPdz}
\end{equation}

\noindent where $\kappa$ is the opacity of the accreting material which
is assumed to be comparable to that of dust, $\kappa_d$.  The vertical gravitational 
acceleration, $g_{z}$ is found from:

\begin{equation}
g_{z}=\Omega^{2}z\mbox{.}\label{eq:gz}
\end{equation}

\noindent The disk, throughout this paper, is assumed to be Keplerian, $\Omega=\Omega_{K}$, where

\begin{equation}
\Omega_{K}=(G M\,R^{-3})^{1/2}\mbox{.}\label{eq:OmegaKepler}
\end{equation}

In the case when  
radiation pressure dominates, the characteristic scale-height of the disk follows from \eqref{eq:dPdz} after neglecting the contribution from the gas
pressure: 

\begin{equation}
H/R\simeq\frac{3}{2}\frac{\mdt{}}{\mdtcr}
\simeq8\times10^{-3}\kappa_{10}\,\mdt[0.1]\,r_{0.01}^{-1}\mbox{,}\label{eq:H2R_rad}
\end{equation}

\noindent where $\mdtcr$ is the Eddington accretion rate, calculated using 
dust opacity $\kappa_{d}$. 

\begin{equation}
\mdtcr=\frac{4\pi cr}{\kappa_{d}}\simeq12.3\,\kappa_{10}^{-1}
\,r_{0.01}\,M_{\odot}{\rm yr}^{-1}\mbox{,}\label{eq:MdotCritDust}
\end{equation}

\noindent where $\kappa_x=\kappa/x$ is the opacity scaled in $x({\rm cm^2\,g^{-1}})$ units of opacity, 
and $\mdt[0.1]$ is the local accretion rate scaled
in $0.1 M_{\odot}{\rm yr}^{-1}$. In a
model of the dust opacity which takes into account the contribution from large graphite grains (BL18) 
the critical mass-accretion rate would be correspondingly smaller:
$\mdtcr = 2.5\,\kappa_{50}^{-1} M_{\odot}{\rm yr}^{-1}$. If the local accretion rate exceeds $\mdtcr$ then the disk scale height becomes comparable to the radius.
For the same set of 
parameters the critical value of $\dot{M}$ calculated assuming only electron scattering
opacity is $\dot{M}_{\text{E}}\simeq 0.2M_{\odot}{\rm yr}^{-1}$ . For the disk not to be super-Eddington globally, the excess mass of the
gas $\sim {\rm few}\MsolYrM$ should be expelled via winds along the way towards the BH.


One can notice that
the gas density in the disk mid-plane is much higher than required by 
(\ref{eq:densPgas=Prad}), so when internal radiation pressure
of the disk is negligible the gas pressure from the disk 
can balance radiation pressure due to external radiation at plausible levels.
If the thick disk is supported by the radiation pressure, the condition
$H/R\propto1$ can be recast as a consequence of the Virial theorem
for the disk temperature, namely $T\simeq T_{{\rm vir,r}}$ where
$T_{{\rm vir,r}}$ is the ``virial'' temperature for the radiation
dominated medium \citep{Dorodnitsyn11a}:

\begin{equation}
T_{{\rm vir,r}}=\left(\frac{GM}{a R}\right)^{1/4}
\simeq 1755.93 \left(\frac{M_{7}(n/10^7)}{r_{0.1}}\right)^{1/4}{\rm \,K}\mbox{,}
\label{eq:TvirRad}
\end{equation}

\noindent derived for a spherically symmetric shell.

\section{Dust Sublimation Region in a Disk}\label{sec:DustSubRegIndisk}

\subsection{The Inner and Outer Dust Sublimation Scales}

In order to calculate the structure of a thin disk,  the
assumption was made in (SS73) that viscous dissipation is proportional
to the gas density $\rho$.  Here we assume that all the dissipated energy is transported vertically
by radiation. This gives the following estimate for the temperature
at the disk surface:

\begin{equation}
T_s=879\;\left(\frac{M_{7}\mdt[0.1]}{r_{0.01}^{3}}\right)^{1/4}\text{\,K}\label{eq:Teff}
\end{equation}

\noindent where it was assumed that $T_{s}=T(\tau_{{\rm phot}}=2/3)$ in which
$\tau_{{\rm phot}}$ is the optical depth at the disk photosphere.
When $R=10^{-3}{\rm pc}$ the surface temperature reaches 
$1479$K and little dust can survive in the disk.

If external illumination is neglected, 
the vertical decrease of temperature
within a disk ensures that the mid-plane temperature, $T_c$
is always greater than the surface temperature $T_{s}$. Approximately, 
$T_{c}$ is a factor of $\tau_{c}^{1/4}$ greater than $T_{s}$
, where $\tau_{c}$ is the vertical optical depth of the disk (see Section \ref{sec:Solution}). 
In general $\tau_{c}$ should be calculated from the solution for
the vertical structure of the disk. From standard gas pressure only
$\alpha-$disk solution we get

\begin{eqnarray}
T_c &=& \frac{1}{2}\left(\frac{3}{2}\right)^{1/5}\kappa^{1/5}J(R)^{2/5}\mu_m^{1/5}
\text{\ensuremath{ \mdt }}^{2/5}\pi^{-\frac{2}{5}}\Omega^{3/5}{\cal R}^{-1/5}
\alpha^{-\frac{1}{5}}\su{\sigma}{B}^{-\frac{1}{5}}\nonumber\\
&\simeq& 2\times10^{3}\kappa_{10}^{1/5}M_{7}^{3/10}
\text{\ensuremath{\mdt[0.1]}}^{2/5}r_{0.1}^{-9/10}{\rm K}\mbox{,}\label{eq:TcAnalytPgas}
\end{eqnarray}
where the factor $J(R)$ is related to the inner boundary condition near the BH and for sub-parsec distances
it is $J\simeq1$ to a good accuracy, and $\su{\sigma}{B}= ac/4$ is the Stefan-Boltzmann constant.

Figure \ref{fig:TSurfVsTcVsTsubSketch} shows graphs of
$T_s(R)$ from \eqref{eq:Teff} and $T_c(R)$ from 
\eqref{eq:TcAnalytPgas}. Two intersections of these two curves with the
$\Tsub$ line define the inner, $\su{R}{in}$ and outer, $\su{R}{out}$ dust sublimation
radii in a disk.
The inner sublimation radius
is defined as the radius where the disk surface temperature equals 
dust sublimation temperature:

\begin{eqnarray}
\su{R}{in} &\simeq& 5\times10^{-3}M_{7}^{1/3}\mdt[0.1]^{1/3}\su{T}{sub,1500}
^{-4/3}{\rm pc}\nonumber\\
&\simeq& 5\times10^{3}M_{7}^{-2/3}\mdt[0.1]^{1/3}\su{T}{sub,1500}^{-4/3}\,r_g.\label{eq:Rin}
\end{eqnarray}

\noindent where $\su{T}{sub,1500}$ is the dust sublimation temperature in units of 1500K.  
Similarly, from (\ref{eq:TcAnalytPgas}), we define an outer sublimation radius, $R_{{\rm out}}$
as such a radius that at $R\le \Rout$ dust sublimes at the mid-plane:

\begin{figure} 
\includegraphics{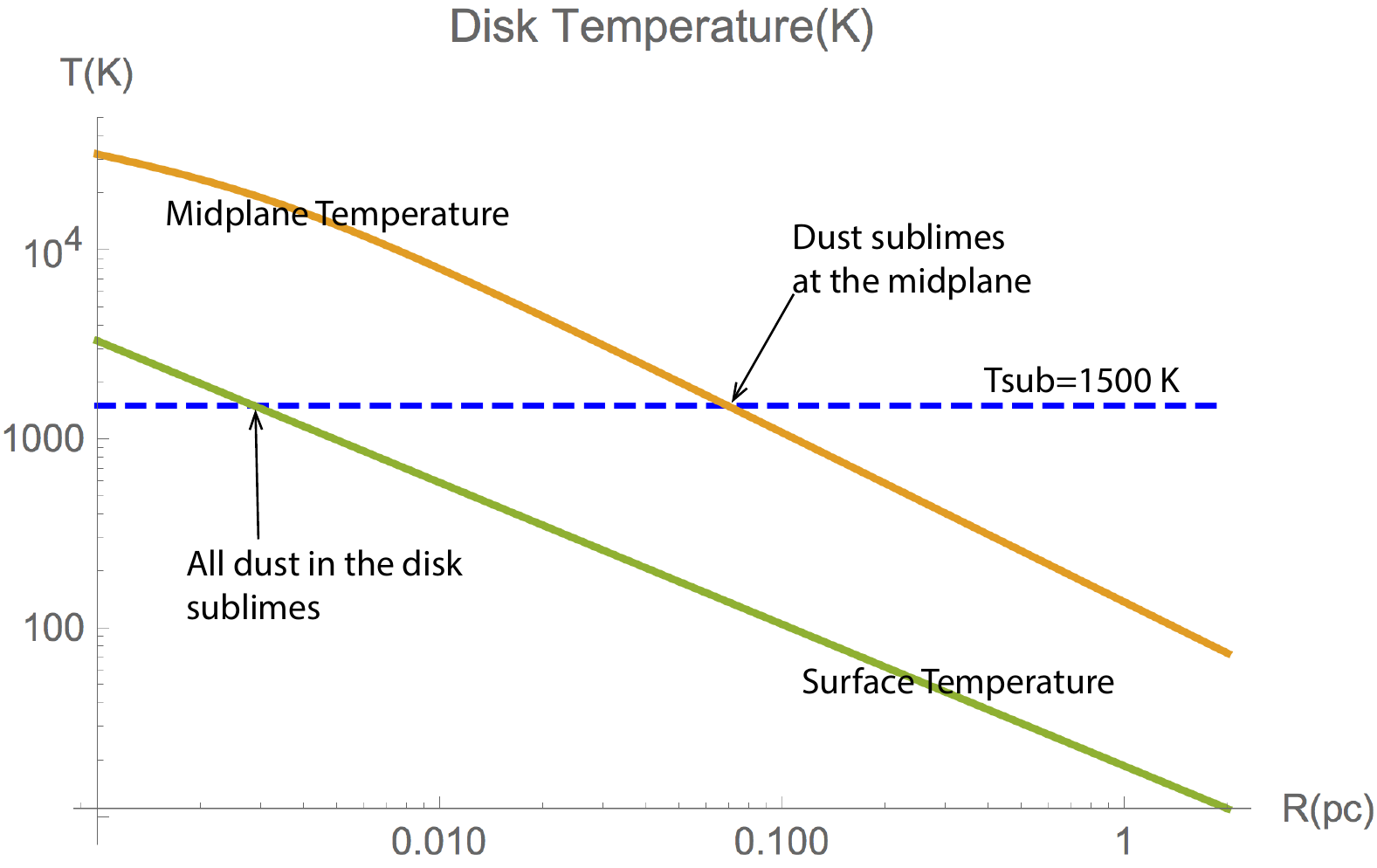}
\caption{Schematics of the dusty region in a disk}
\label{fig:TSurfVsTcVsTsubSketch}
\end{figure}

\begin{eqnarray}
\su{R}{out}  &\simeq&  0.14M_{7}^{1/3}\kappa_{10}^{2/9}
\mdt[0.1]^{4/9}
\alpha_{0.1}{}^{-\frac{2}{9}}\su{T}{sub,1500}^{-10/9}{\rm pc}\nonumber\\
&\simeq&
1.5\times10^{5}\mdt[0.1]^{4/9}\kappa_{10}^{2/9}M_{7}{}^
{-\frac{2}{3}}\alpha_{0.1}{}^{-\frac{2}{9}}\su{T}{sub,1500}^{-10/9}\,r_g\mbox{.}\label{eq:Rout}
\end{eqnarray}

\noindent Since $\su{R}{out}$ is found from $T(\Sigma_c) = \Tsub$, 
the disk's column density, $\Sigma_c = 2\rho_c H$  
should found from the solution for the disk.

As the gas spirals to $R\le \Rout$, $\Tc$ becomes greater
than dust sublimation temperature $T_{{\rm sub}}$. Dust is destroyed
first near the mid-plane and then, as $r$ gets smaller, progressively
above. Naturally $R_{{\rm out}}>R_{{\rm in}}$ defining the region
within the disk where hot dust exists at some height in the disk. Notice that at $R=R_{{\rm out}}$
disk surface is quite cool: $T_{{\rm s}}\simeq156\left(M_{7}\dot{M}_{0.1}\right){}^{1/4}r_{0.1}{}^{-3/4}{\rm K}$.
Such a difference between $T_s$ and $T_c$ is essentially the 
result of the blanketing effect from the disk.
Dependence of $\su{R}{in}$ and $\su{R}{out}$ on $M$ follows 
if we assume the scaling $\mdt\propto\Mdt\propto \su{L}{Edd} \propto M$:

\begin{equation}
\su{R}{in}\propto M^{2/3} \mbox{,}
\end{equation}
and
\begin{equation}
\su{R}{out}\propto M^{7/9}\mbox{.}
\end{equation}

\noindent Increasing the mass of the BH, we obtain $R_{{\rm in}}\simeq0.1$pc , and $R_{{\rm out}}\simeq5$pc 
for $M=10^{9}M_{\odot}$.

The accretion rate near the BH, $\Mdt$ determines the central luminosity  $L$ through \eqref{eq:AGNLum1} and thus the global dust sublimation radius $\rsub$ 
(see Section \ref{sec:DustSubRegIndisk}). 

\begin{equation}\label{eq:RsubAGN}
  \su{R}{sub}=0.13\left(\frac{f(\theta)\epsilon_{0.1}{\dot{M}}_{0.1}}{T_{1500}^{4}}\right)^{1/2}\,\text{pc}\mbox{,}
\end{equation}

\noindent where 
$f(\theta)=\mu(2\mu+1)$ is the angular dependence of the radiation flux from the nucleus, i.e. \eqref{eq:FextFormula}. 
The result is the dust sublimation surface, which, generally speaking, is different from a simple spherically symmetric case such as 
\eqref{eq:RsubAGN} which was calculated for $f=1$. 
Notice that there is a difference between our definition of $\su{R}{out}$
in \eqref{eq:Rout} and the definition of BL18. The latter
define $\su{R}{out}$ as the dust sublimation radius for AGN
\eqref{eq:RsubAGN} because their work studies the size of the BLR rather than the structure and properties
of the disk itself. 
As long as energy is transported vertically via radiation, it follows
from (\ref{eq:Rout}) and \eqref{eq:Rin} that $\Rout\propto\mdt[0.1]^{1/9}\Rin$.
Another interesting scaling: $\su{R}{sub}/R_{{\rm out}}\sim (\Mdt/\mdt)^{1/2}$ 
follows from \eqref{eq:Rout}, \eqref{eq:RsubAGN}.

\begin{figure}
   \includegraphics[scale=0.7]{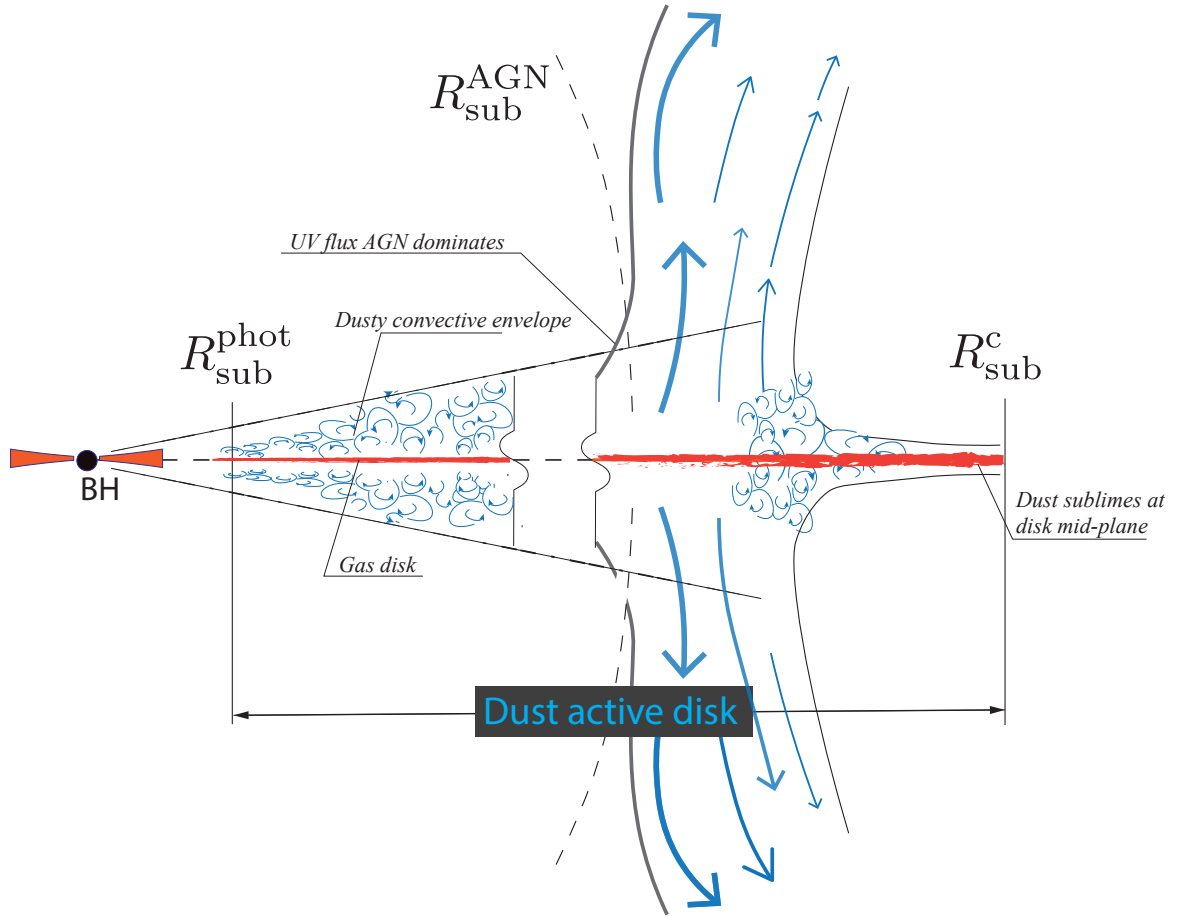} 
  \caption{\label{fig:DiskSketch} 
  Convective dusty disk with an outflow.
  When $\mdt\gtrsim \Mdt  (L_{{\rm bol}})$ (i.e accretion is locally supercritical)
  disk's own radiation pressure pushes dust above the disk. 
  At  $R>\su{R}{sub}^{\rm AGN}$ disk outflow is launched by the disk's 
  own IR pressure, with a possibility to being further accelerated by the radiation
  pressure from the nucleus.
  Not to scale.}
  \end{figure}

Without the shielding protection of the accretion disk the fate of
dust above such disk depends on whether it is inside or outside the
AGN dust sublimation surface $\rsub(\theta)$:

\begin{enumerate}
  \item if $\rsub \gtrsim \su{R}{out}$, the dust above the disk does not survive.
  \item if $\su{R}{in} \lesssim \rsub \lesssim \su{R}{out}$, depending on $\mdt$ the disk can 
  be 1) thin or 2)thick/outflowing, (the situation is illustrated in
  Figure \ref{fig:DiskSketch}).  
\end{enumerate}

When $\mdt$ exceeds 
\begin{equation}
    \su{\dot{M}}{loc}(\rsub =  \su{R}{out} ) = 0.08 \left(\frac{ M_7^{1/3} T_{1500}^{8/9} \kappa_{10}^{2/9}}
    {\alpha_{0.1}^{2/9} \epsilon_{0.1}^{1/2}}\right)^{-9/4} 
    \dot{M}_{0.1}^{9/8}\,\MsolYrM\mbox{,}
\end{equation}
the condition $\rsub \le  \su{R}{out}$ is fulfilled, at least part of the active region 
above the disk is shielded beyond the AGN dust sublimation surface. 

\section{Solution for the Disk Vertical Structure}\label{sec:Solution}
\subsection{Basic equations}

In this section we describe the details of our model used to derive estimates 
in the previous sections. 
We adopt the $\alpha-$disk theory of SS73 and include pressure of infrared (IR) radiation on dust in the disk 
interior. The radiation is produced by viscous dissipation and assumed to be in thermodynamic 
equilibrium with gas-dust mixture. Dust particles are assumed to be fully coupled to the gas. 

The equation of state is that of a mixture of ideal gas and radiation:

\begin{equation}
P=P_g+P_r\mbox{,}\label{eq:Ptot}
\end{equation}
where $P$ is the total pressure, $P_r$ is the radiation pressure defined 
in \eqref{eq:Prad}, and  $P_{g}$ is the gas pressure defined in \eqref{eq:Pgas}.
Analytic solutions for the
limiting cases of $P_g \ll P_r$, and $P_g \gg P_r$ are derived in
\citep{ShakuraSunyaev73}. Here we will solve the disk equations for arbitrary
$P_g$ and $P_r$ assuming constant opacity.

We assume that in the ADR region, accretion proceeds near the equatorial plane:

\begin{equation}
\dot{M}=2\pi r v\,\Sigma\mbox{,}\label{eq:MdotEq}
\end{equation}

\noindent where to simplify notation in this section $\dot{M}=\mdt$ is the time averaged accretion rate, $v$ is the radial
gas velocity and $\Sigma$ is the surface density: 

\begin{equation}
\Sigma=\int_{-H}^H\rho\,dz\simeq2 H\,\rho\mbox{,}\label{eq:SigmaS}
\end{equation}

\noindent where $H$ is the vertical scale-height of the disk and $\rho=\rho(z=0)$,
i.e. in a single-layer disk model, all height-dependent quantities are taken at the
disk mid-plane.

\begin{equation}
\nu\Sigma=\frac{1}{3\pi}\dot{M}\mbox{,}\label{eq:AngMom}
\end{equation}

\noindent where we neglected a factor, related to the inner boundary condition,
(i.e. $J(R)$ in \eqref{eq:TcAnalytPgas}).
The viscosity $\nu$ in \eqref{eq:Nu_HTSigma} is proportional to the total pressure, $P$:

\begin{equation}
  \nu=\alpha\,\frac{P}{\text{\ensuremath{\Omega\rho}}}\mbox{.}
  \label{eq:defNuAlphaPtot}    
\end{equation}

\noindent The only relevant component of the radiation flux in a pc-scale thin disk
is the vertical one. 
The vertical flux $F$ satisfies: 

\begin{equation}
\frac{\partial F}{\partial z}=\frac{9}{4}\rho_c
\,q_{{\rm v}}\equiv\frac{9}{4}\Omega^{2}\rho_c\nu\mbox{,}\label{eq:dFdz}
\end{equation}

\noindent where $q_{{\rm v}}$$({\rm ergs\cdot s^{-1}\cdot g^{-1}})$ is the
specific rate of viscous energy dissipation: 

\begin{equation}
\su{q}{v}=\Omega^{2}\nu\mbox{.}\label{eq:qv(specific)}
\end{equation}

\noindent Integrating equation (\ref{eq:dFdz}) between $-H$ and
$+H$, one gets the vertical radiation flux from the disk surface:

\begin{equation}
F^{+}=\frac{3}{8\pi}\dot{M}\Omega^{2}\mbox{.}\label{eq:RadFlux}
\end{equation}

\noindent An implicit assumption was made when integrating \eqref{eq:dFdz} to obtain \eqref{eq:RadFlux}. 
That is, after equation \eqref{eq:dFdz} was rewritten as
${\displaystyle \frac{\partial F}{\partial\sigma}=\frac{9}{4}q_{{\rm v}}}$, where 

\begin{equation}\label{eq:MassCoordinate}
\sigma(z)=\int_0^z \rho\,dz \mbox{,}
\end{equation}
is the mass coordinate, and then integrated over height, 
and it was assumed that the rate of \emph{specific} 
viscous dissipation, $\su{q}{v}$ is constant. 

The radiation moment equation in
a plane-parallel case valid for the geometrically thin disk is:

\begin{equation}
\frac{\partial E}{\partial\tau}=3\frac{F}{c}\mbox{,}\label{eq:dEdTau}
\end{equation}

\noindent where $E$ and $F$ are radiation energy density and radiation flux
and $\tau$ is the vertical optical depth. 

The boundary condition for \eqref{eq:dEdTau} is $E(\tau=0)=2F^+/c$. 
Integrating \eqref{eq:dEdTau} results in $E=(3F^+/c)(\tau+2/3)$ and

\begin{equation}\label{eq:SigmaT4c}
  \su{\sigma}{B}T^4 = \frac{27}{64}\nu\kappa\,\Omega^2\Sigma^2\mbox{,} 
\end{equation} 

\noindent introducing the subscript "c" for mid-plane quantities, 
from (\ref{eq:SigmaT4c}) it follows that if $\tau_{c}\gg1$, 
where 
${\displaystyle
\tau_c = \int_{0}^{H}\kappa \, \rho \,dz\simeq \Sigma \, \kappa_d }$
then (\ref{eq:TcAnalytPgas}) gives: 

\begin{equation}
T_c\propto\tau_{c}^{1/4}T_{{\rm s}}\mbox{,}\label{eq:TcTauc}
\end{equation}

\subsection{Solution with radiation pressure }\label{sec:SolWithRadPressure}
The solution for $\Tc$ for gas-dominated disk
was given in (\ref{eq:TcAnalytPgas}).
{\mybf As $T$ approaches $\Tsub$, the radiation pressure becomes important. When $T_c>\Tsub$ a gas-only layer at the mid-plane is enveloped in 
a dusty-gaseous envelope above the mid-plane. From above its temperature is bound by $T^{+}\lesssim \Tsub$, and the transition from gas-dominated layer to dust-dominated envelope is happening at the height $H_g$. The estimate of scale-height $H_g$ follows from \eqref{eq:dPdz}:

\begin{eqnarray}
\label{Hgas}
H_g/R &\simeq& \left(\frac{P_c-P^{+}}{\rho_c \Omega^2 }\right)^{1/2}
\lesssim  \left(\frac{   {\cal R} \Tsub  } 
{\Omega^2}\right)^{1/2}\nonumber\\
&\simeq& 5.4 \times 10^{-3} (\su{T}{1500} r_{0.1}/M_7 ) ^{1/2}  \mbox{,}
\end{eqnarray}
i.e. the gas layer is very geometrically thin.

It is beyond of the scope of this paper to calculate the vertical structure of the disk with the gas to dust transition. Instead, we adopt a single layer approximation with the
temperature dependent opacity (see further \eqref{eq:kappaLogistic})
}
Thus in the following our goal is to
calculate the {\it average} properties of the disk which is 
supported by an arbitrary combination of gas and radiation pressure: $P=P_{{\rm gas}}+aT^{4}/3$. 

Assuming the relation:

\begin{equation}
{\displaystyle \rho=\frac{\Sigma}{2H}}\label{eq:rhoEqSigmaOver2H}\mbox{,}
\end{equation}
then the total pressure, $P$ (reminding that the molecular weight is absorbed in
the gas constant ${\cal R}$) is

\begin{equation}
P=\rho\left({\cal R}T+\frac{2aT^{4}H}{3\Sigma}\right)\mbox{.}
\label{eq:nuAlphaPtot}
\end{equation}

\noindent The approximate relation for the scale-height follows from
the equation for the vertical balance:

\begin{equation}\label{eq:VertBalaceTotalPNoFlux}
\displaystyle \frac{dP}{dz}=-\rho\,g_{z},    
\end{equation}
which is equivalent to \eqref{eq:dPdz}, after noticing that in diffusion approximation 

\begin{equation}\label{eq:FlocDiffusApprox}
\displaystyle F=-c/(\kappa\rho)dP_r/dz\mbox{,}
\end{equation}
{ and thus recovering:}
  
\begin{equation}
H^{2}=\frac{P}{\Omega^{2}\rho}\mbox{.}\label{eq:ScaleHeight}
\end{equation}
{ Equation \eqref{eq:ScaleHeight} is the equivalent of the vertical balance equation.
Substituting \eqref{eq:nuAlphaPtot} and \eqref{eq:rhoEqSigmaOver2H} to \eqref{eq:ScaleHeight} there follows a 
useful relation:
}
\begin{equation}\label{eq:EqSigmaFromHT}
  \Sigma = \frac{2 a H T^4}{3 \left(H^2 \Omega ^2-{\cal R} T\right)}\mbox{.}  
\end{equation}

\noindent
{\mybf
After substituting \eqref{eq:nuAlphaPtot} and \eqref{eq:rhoEqSigmaOver2H} 
to \eqref{eq:defNuAlphaPtot} there follows
another relation:}

\begin{equation}\label{eq:Nu_HTSigma}
\nu=\frac{\alpha  \left(2 a H T^4+3 {\cal R} \Sigma  
T\right)}{3 \Sigma  \Omega }\mbox{.}
\end{equation}

{\mybf The third useful relation we can get from a system of two equations for 
$\Sigma$ and $H$ (and $T$)
which is obtained by first inserting  
$\nu$ from \eqref{eq:SigmaT4c} into \eqref{eq:AngMom}
and then repeating by inserting $\nu$ 
from  \eqref{eq:Nu_HTSigma} to the 
\eqref{eq:AngMom} and then solving these two equations for $H$:
}
\begin{equation}\label{eq:Height_FT}
H= \frac{4 F^+}{3 a \alpha  T^4 \Omega }-\frac{c {\cal R} T}{F^+ \kappa }\mbox{.}
\end{equation}

{\mybf We now can use \eqref{eq:nuAlphaPtot}, \eqref{eq:rhoEqSigmaOver2H} with
\eqref{eq:ScaleHeight}, \eqref{eq:EqSigmaFromHT},
\eqref{eq:Nu_HTSigma}, and  \eqref{eq:Height_FT} and after much algebra we 
obtain an equation for
$T$:

\begin{equation}\label{eq:TcNotReduced}
  4 \left(F^+\right)^2 \kappa  \left(6 a \alpha  c^2 {\cal R} 
  T^5 \Omega ^2+\left(F^+\right)^2 \kappa  \left(3 a \alpha  \kappa  T^4-4 c \Omega \right)\right)
  -9 a^2 \alpha ^2 c^3 {\cal R}^2 T^{10} \Omega ^3=0 \mbox{.}
\end{equation}
Equation \eqref{eq:TcNotReduced} can be further simplified down to an equivalently rather
compact form:
\begin{equation}
  \left(1-\frac{3a\alpha c{\cal R}\Omega}{4\left(F^{+}\right)^{2}\kappa}
  \Tc^{5}\right)^{2}-\frac{3a\alpha\kappa}{4c\Omega}\Tc^{4}=0\mbox{.}
  \label{eq:TcNonLinEq1}
 \end{equation}
Thus the solution for the disk temperature should be calculated from 
the non-linear algebraic equation \eqref{eq:TcNonLinEq1}.
It can be shown that if radiation pressure is neglected, term in the brackets
in \eqref{eq:TcNonLinEq1} is left and the corresponding expression for
gas pressure alpha disk is recovered:

\begin{equation}
  \su{T}{c}(P_g) = \frac{(3 \kappa)^{1/5} (\dot{M})^{2/5} \Omega^{3/5}}
  { 2^{4/5} \pi^{2/5} (\alpha a c  R)^{1/5}
  }\mbox{,}
  \label{eq:TdiskGasFromNonLinSimplification}
\end{equation}
from which, for example, the numerical 
scaling \eqref{eq:TcAnalytPgas} can be calculated.
}

\subsection{Numerical solution}\label{sec:NumericalSolution}
At a given $r$, with $F^{+}$ calculated from (\ref{eq:RadFlux})
and with $\kappa$ from (\ref{eq:kappaLogistic}) and $\Omega$ from
(\ref{eq:OmegaKepler}), equation (\ref{eq:TcNonLinEq1}) is solved
numerically. However, some tricks are required 
to ensure numerical stability. First 
we assume that above $T_{{\rm sub}}$ the opacity switches from $\kappa_{d}+\kappa_{e}$ to
$\kappa_{e}$  and approximate $\kappa(T)$ by a bridging formula:

\begin{equation}
\kappa(T)=\frac{\kappa_d}
{\exp\left(
  \frac{T-\su{T}{sub}}{\Delta T}
  \right)+1} + \kappa_e
\mbox{,}
\label{eq:kappaLogistic}
\end{equation}

\noindent where the bridging parameter $\Delta T$ is fixed at 
$\Delta T=0.1$ as well as $\kappa_{d}=10\,\text{cm}^{2}\text{g}^{-1}$.
Then, to find roots of equation \eqref{eq:TcNonLinEq1} the following
procedure is adopted:
initial approximation for $T_0$
is found after $\kappa(T_0)$ is initially estimated from (\ref{eq:kappaLogistic}), 
where $T_0$ obtained
for the gas-pressure only solution \eqref{eq:TcAnalytPgas}.
Then $T_0$ is used as initial guess to numerically 
solve (\ref{eq:TcNonLinEq1}) with (\ref{eq:kappaLogistic}) 
for the true value of $T$. Finally,
multiple roots of equation \eqref{eq:TcNonLinEq1} should be weeded out via checking that
they produce positive right-hand-side in the equation (\ref{eq:ScaleHeight}).

Once $\Tc$ is known the surface
density $\Sigma$ is found form equations 
(\ref{eq:AngMom}),\eqref{eq:nuAlphaPtot} and (\ref{eq:Ptot}),
and (\ref{eq:rhoEqSigmaOver2H}):

\begin{equation}
\Sigma=\frac{64\pi\sigma}{9 \kappa} 
\frac{\Tc^{4}}{\dot{M}\Omega^{2}}
\mbox{.}\label{eq:SigmaSolArbP}
\end{equation}

\begin{figure}
  \includegraphics[scale=0.7]{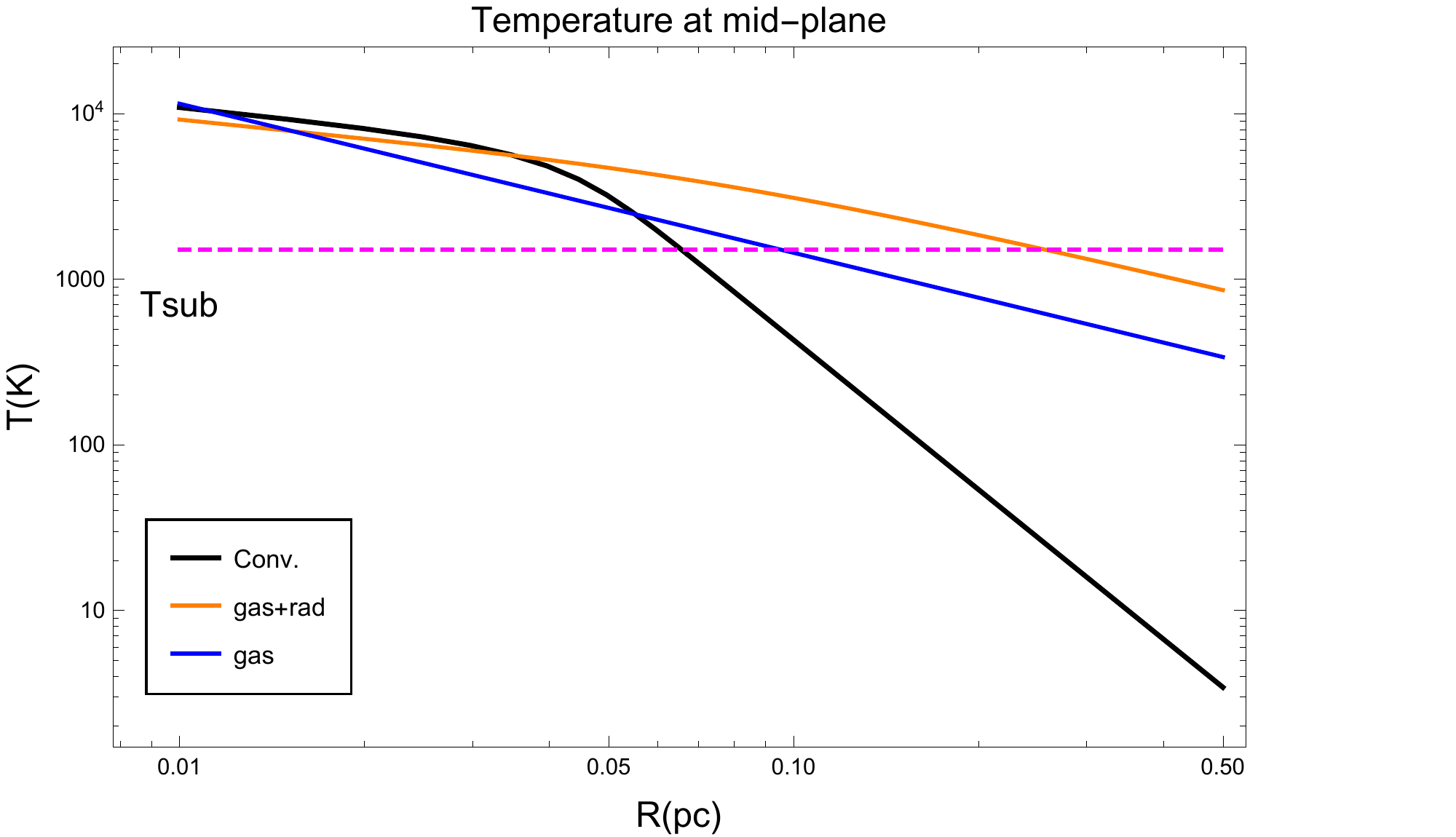} 
  \caption{
  Disk temperature at the disk mid-plane.  
  Black: convective.  Blue: gas+radiation pressure, 
  equation \eqref{eq:TcNonLinEq1}.
  Orange: gas-pressure-only,
  equation \eqref{eq:TcAnalytPgas}. 
  The intersection with the dashed line, $T_{{\rm sub}}=1500$K marks the 
  position of $\Rout$.}
  \label{fig:RadDiskTempPlot} 
\end{figure}

\noindent The result of a numerical solution of the equation 
\eqref{eq:TcNonLinEq1}
with respect to $\Tc$ is shown in 
Figure (\ref{fig:RadDiskTempPlot}) where 
disk model for $P=P_g$ is compared with the model for
$P=P_g+P_r$.

While this figure more accurately illustrates the relation between
$\Rin$ and $\Rout$, we still probably underestimate $\Rout$
because $\kappa_d$ becomes important before $T$ is approaching
$T_{{\rm sub}}$. Thus, the region where the dust opacity changes the
vertical structure of the disk is extended further away. From
\eqref{eq:Rin} and \eqref{eq:Rout} one can see that the size of this
region relatively weakly depends on $\dot{M}$.
{\mybf Solving equation \eqref{eq:TcNonLinEq1} for different values of
parameters, we calculate the size of the active region, i.e.
$\Rin$ and $\Rout$. Table \ref{table:1} summarizes the results for 
the dependence of $\Rin$ and $\Rout$ on 
$\mdt$ and column for $\su{M}{BH}$.
Each pair is shown  at the intersection of the corresponding 
row for $\mdt$ and column for $\su{M}{BH}$.}

\newcommand{\Rie}[1]{\su{R}{in}  = #1}
\newcommand{\Rio}[1]{\su{R}{out} = #1}

\begin{table}[h!]
\centering
\begin{tabular}{|c|c|c|c|c|c|}
  \hline
  {\diagbox{$\frac{\mdt}{M_{\odot}\text{yr}^{-1}}$}
  {$\frac{\su{M}{BH}}{M_\odot}$}}
  & $10^{5}$ &  $10^{6}$ 
  & $10^{7}$ & $10^{8}$ & $10^{9}$\\
  \hline
      0.01 & $5.53\times 10^{-3}$ & $7.29\times 10^{-3}$ 
       & $8.61\times 10^{-3}$ & $9.57\times 10^{-3}$ & $1.96\times 10^{-2}$\\ 
      0.01 & $1.09\times 10^{-2}$ & $1.81\times10^{-2}$ 
      & $7.17\times 10^{-2}$ & $1.54\times 10^{-1}$ & $3.33\times 10^{-1}$\\ 
  \hline
      0.1 & $7.29\times 10^{-3}$ & $8.61\times 10^{-3}$ 
      & $9.57\times 10^{-3}$ & $1.02\times 10^{-2}$ & $2.04\times 10^{-2}$\\ 
      0.1 & $4.15\times 10^{-2}$ & $8.94\times 10^{-2}$ 
      & $1.93\times 10^{-1}$ & $4.15\times 10^{-1}$ & $8.94\times 10^{-1}$\\ 
  \hline            
  1 & $8.61\times 10^{-3}$ & $9.57\times 10^{-3}$ 
    & $1.02\times 10^{-2}$ & $2.25\times 10^{-2}$ & $4.83\times 10^{-2}$\\ 
  1 & $1.06\times 10^{-1}$ & $2.28\times 10^{-1}$ 
    & $4.9\times 10^{-1}$ & $1.06$  & $2.28$  \\
  \hline            
    2 & $1.72\times 10^{-2}$ & $9.79\times 10^{-3}$ 
      & $1.04\times 10^{-2}$ & $2.84\times 10^{-2}$ & $6.18\times 10^{-2}$\\ 
    2 & $1.37\times 10^{-1}$ & $2.94\times 10^{-1}$ 
               & $6.34\times 10^{-1}$ & $1.36$ & $2.91$\\
  \hline            
    10 & $9.57\times 10^{-3}$ & $1.02\times 10^{-2}$ 
     & $2.25\times 10^{-2}$ & $4.91\times 10^{-2}$ & $1.06\times 10^{-1}$\\ 
    10 & $2.21\times 10^{-1}$ & $4.76\times 10^{-1}$ 
    & $1.03$ & $1.86$ & $3.83$\\             
  \hline
  \end{tabular}
  \caption{$\Rin$ and $\Rout$ calculated from the numerical solution 
  of equation \eqref{eq:TcNonLinEq1}. Left column: accretion rate, 
  $\mdt$. Upper raw: mass of the Black Hole, $\su{M}{BH}$. 
  Each cell not belonging to the first raw or the first column
  contains two numbers: upper number: $\Rin$; lower number: $\Rout$.
  }
  \label{table:1}
\end{table}

\section{Convection}\label{sec:Convection}

If a gas element which is rising over a small distance, adiabatically
and in pressure equilibrium with the environment, is found to 
be lighter when contrasted to its surroundings, then
buoyancy force will keep propelling it further \citep{KippenhahnWeigert94}.
The opposite situation corresponds to the  Schwarzschild criterion for convective
stability:

\begin{equation}
\left|\frac{dT}{dz}\right|_{{\rm ad}}>\left|\frac{dT}{dz}\right|\mbox{.}\label{eq:SchwarzschildCrit}
\end{equation}

Where $\left(\frac{dT}{dz}\right)_{{\rm ad}}$ is the adiabatic and 
$\frac{dT}{dz}=\left(\frac{dT}{dz}\right)_{{\rm rad}}$ radiative temperature gradients (both negative).
In this section we revisit assumptions about vertical transport of
energy in the disk 
and show, that as soon as the mid-plane 
temperature exceeds the sublimation temperature, $\Tc \sim T_{{\rm sub}}$, there is a certain range of radii between
$\Rin$ and $\Rout$, where the disk is convectively unstable. There are two reasons for
convection: 1) As we will show the regular 
Schwarzschild criterion for the radiative disk indicates that in a wide range of parameters
transferring energy vertically by convection is preferred over radiation diffusion.
2) Strong temperature-dependence of the opacity of dust leads to a sudden increase of the radiation pressure at some height above the midplane.

The first point is analogous to convective instability of a radiation-dominated 
part of a standard $\alpha$-disk. 
In a disk in radiative equilibrium, the entropy, $S_r$, falls abruptly with increasing $z$ and
convection drives the equilibrium towards an isentropic
state: $S_{r}\propto const.$ \citep{BKBlinn77}, where $S_{r}$ is
the entropy of the gas of radiation when $P_r\gg P_{g}$:

\begin{equation}
S_{r}=\frac{4}{3}\frac{aT^{3}}{\rho}\mbox{.}\label{eq:EntropyRad}
\end{equation}

The second point follows from that in a disk in radiative equilibrium 
$T(z)$ decreases from the mid-plane (cleared from dust at $T>\Tsub$) towards higher $z$.
A dramatic increase of the radiation pressure associated
with an opacity jump follows. 
As long as $T_{c}>T_{{\rm sub}}$,
there exists $z_{s}(r)$ within a vertical column of the disk where there
is a transition from dust-free opacity, $\kappa_{m}$ to dust opacity,
$\kappa_{d}$. Correspondingly $|dT/dz|$ becomes very large at $z_{s}$, 
triggering convection, which works towards smoothing the vertical distribution
of the entropy.

The convection establishes a new distribution of $\rho$ and $T$
so as to decrease $|dT/dz|$. 
Our simplified model discussed so far adopts vertically averaged
quantities and a 
more elaborate treatment of convection calls for more
sophisticated methods. The latter is
beyond the scope of this paper and 
in our derivation we  estimate the efficiency of convection
adopting the solution for the disk in radiative equilibrium  as an initial condition.

\subsection{Convective region}\label{sec:ConvectiveRegion}

Our adopted fiducial set of parameters include $R=0.1$ pc, $T_c=\Tsub=1500$K i.e. corresponding to the
situation of the dust at exactly the dust sublimation temperature at the disk mid-plane. 
Estimating $dT/dz$ for
$n = 1.84\times10^{11}T_{1500}^{3}\,\text{cm}^{-3}$
corresponding to $P_{{\rm g}}=P_r$ case \eqref{eq:densPgas=Prad}, we have
 
\begin{equation}
\frac{dT}{dz}\propto\frac{T_{{\rm s}}-T_{c}}{H}\simeq-2.7\times10^{-12}\:{\rm K}\cdot{\rm cm}^{-1}\mbox{,}\label{eq:dTdz_Model_num}
\end{equation}

\noindent where $T_{s}$ and $T_{c}=T(z=0)$ are estimated from the disk model.
This should be compared with the adiabatic gradient 
$\left(\frac{dT}{dz}\right)_{{\rm ad}}$ estimated from:

\begin{eqnarray}
\left(\frac{dT}{dz}\right)_{{\rm ad}}&=&
-g_{z}\rho\left(\frac{\pd T}{\pd P}\right)_{s}=-g_{z}\rho H \nonumber\\
&\simeq& -1.3\times10^{-12}\:{\rm K}\cdot{\rm cm}^{-1}\mbox{\mbox{,}}\label{eq:dTdz_ad_num}
\end{eqnarray}

\noindent where the disk height $H$ is calculated taking into account the equation
of state (\ref{eq:Ptot}). The above estimations are too crude to adopt them in 
a judgement on convective instability, however
from
(\ref{eq:dTdz_Model_num}) and (\ref{eq:dTdz_ad_num}) it follows that
in the region of $\Rin\lesssim r\lesssim \Rout$ adiabatic
$\left(\frac{dT}{dz}\right)_{{\rm ad}}$ can have similar magnitude
as $\frac{dT}{dz}$ warranting further investigation. 

For the mixture of gas and radiation, the convective flux is \citep{KippenhahnWeigert94,BisnovatyiKogan2001book}:

\begin{equation}
F_{{\rm conv}}=\frac{1}{4}l^{2}\rho\,C_{p}\left(\frac{g_{z}}{T}\right)^{1/2}(\Delta\nabla T)^{3/2}\sqrt{\frac{4P_r}{P_{g}}+1}\mbox{,}\label{eq:Fconv}
\end{equation}

\noindent where $l$ is the mixing length, $C_{p}$ is the heat capacity at constant pressure for the mixture
of gas and radiation:

\begin{equation}
C_{p}={\cal R}\left(\left(\frac{4P_r}{P_{g}}\right)^{2}+\frac{20P_r}{P_{g}}+\frac{5}{2}\right)\mbox{.}\label{eq:Cp}
\end{equation}

\noindent The temperature excess of the convective element over its surroundings
is represented by $(\Delta\nabla T)$ which is found from:

\begin{equation}
\Delta\nabla T=\left(\frac{\pd T}{\pd P}\right)_{s}\frac{dP}{dz}-
\frac{dT}{dz}=-g_{z}\rho\left(\frac{\pd T}{\pd P}\right)_{s}-
\frac{dT}{dz}\mbox{.}\label{eq:DelNablaT}
\end{equation}

\begin{figure}
\includegraphics[scale=0.7
]{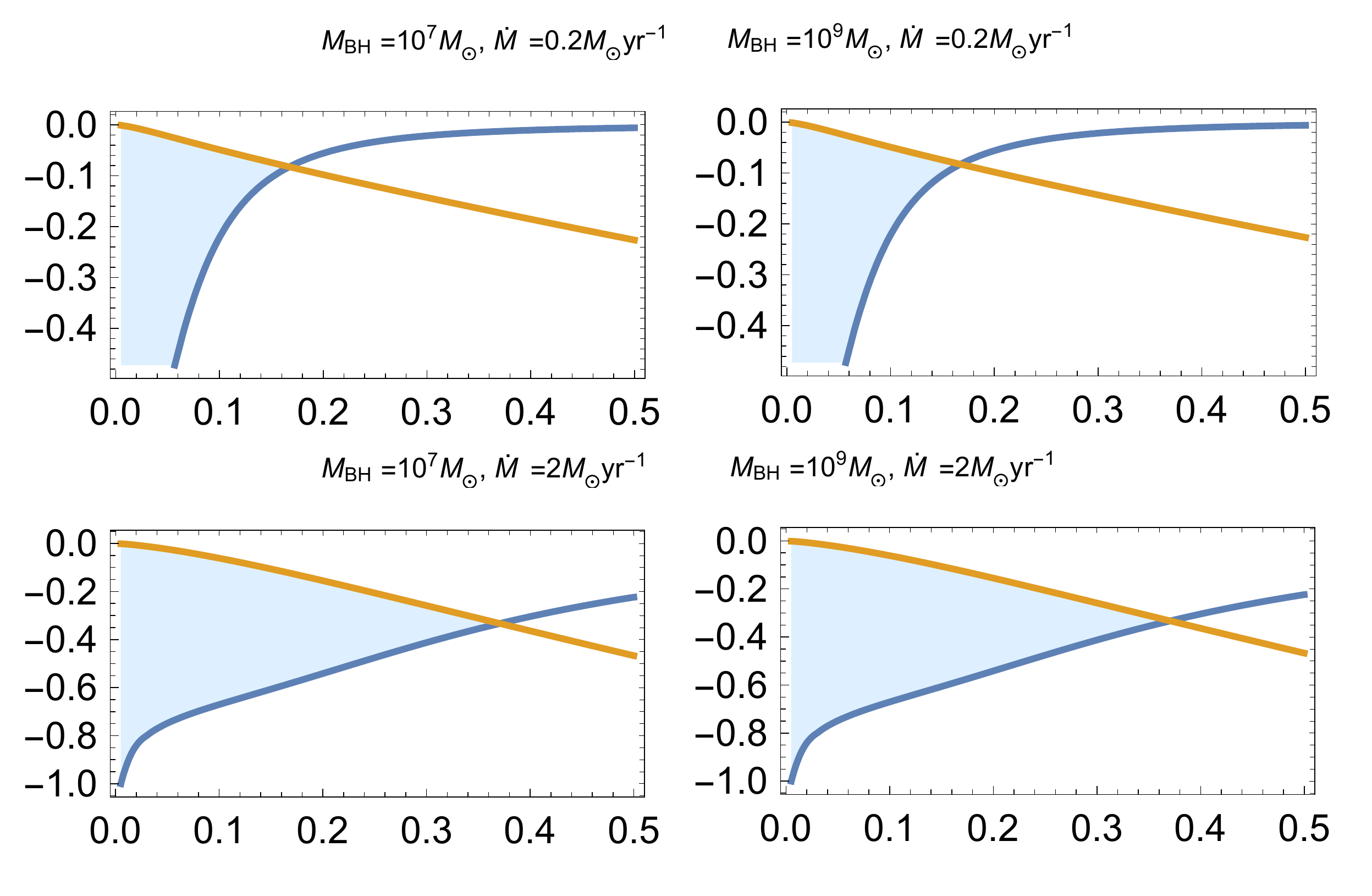} 
\caption{Transition to convection at the disk mid-plane. Shown are non-dimensional
(in code units) gradients: $\left(\frac{dT}{dz}\right)_{{\rm ad}}$
-orange line; actual $\frac{dT}{dz}$ - blue line. Horizontal axis:
distance from the BH in pc. Shaded area: region of convective instability.}
\label{fig:ConvRegion}
\end{figure}

\noindent We adopt $l\simeq\epsilon_{0}H$ for the mixing length, where
$H$ is the half-thickness of the disk and $\epsilon_{0}\leq1$ is
the mixing length parameter for which we adopt the value $\epsilon_{0}=0.1$.
Adopting  $T$ and $\rho$ from the disk model from Section (\ref{sec:SolWithRadPressure}), we numerically 
calculate $\su{F}{conv}$ from \eqref{eq:Fconv} and \eqref{eq:DelNablaT}.
The result is shown in 
Figure \ref{fig:ConvRegion},
where non-dimensional
$\su{\left(\frac{dT}{dz}\right)}{ad}$ and $\left(\frac{dT}{dz}\right)$ are plotted. Consider the situation when $r$ is decreasing
(i.e. tracing the diagram from right-to-left): the curves cross when 
$\su{\left(\frac{dT}{dz}\right)}{ad}>\frac{dT}{dz}$ and 
the medium becomes convectively unstable. 
In the left column the effect of accretion rate is shown: for $\mdt=0.1\MsolYrM$ the ADR is convective at $\Rin<r<0.13$pc;
increasing the accretion rate pushes the convective region further out: for $\mdt=2\MsolYrM$, the convective region is at $\Rin<r<0.3$pc.
In the right column one can see a similar effect if the mass of the BH is increased to $\su{M}{BH}=10^9M_\odot$.

\section{Effects of external irradiation}\label{sec:ExternalIrradiation}

\begin{figure}[h]
\includegraphics[scale=1]{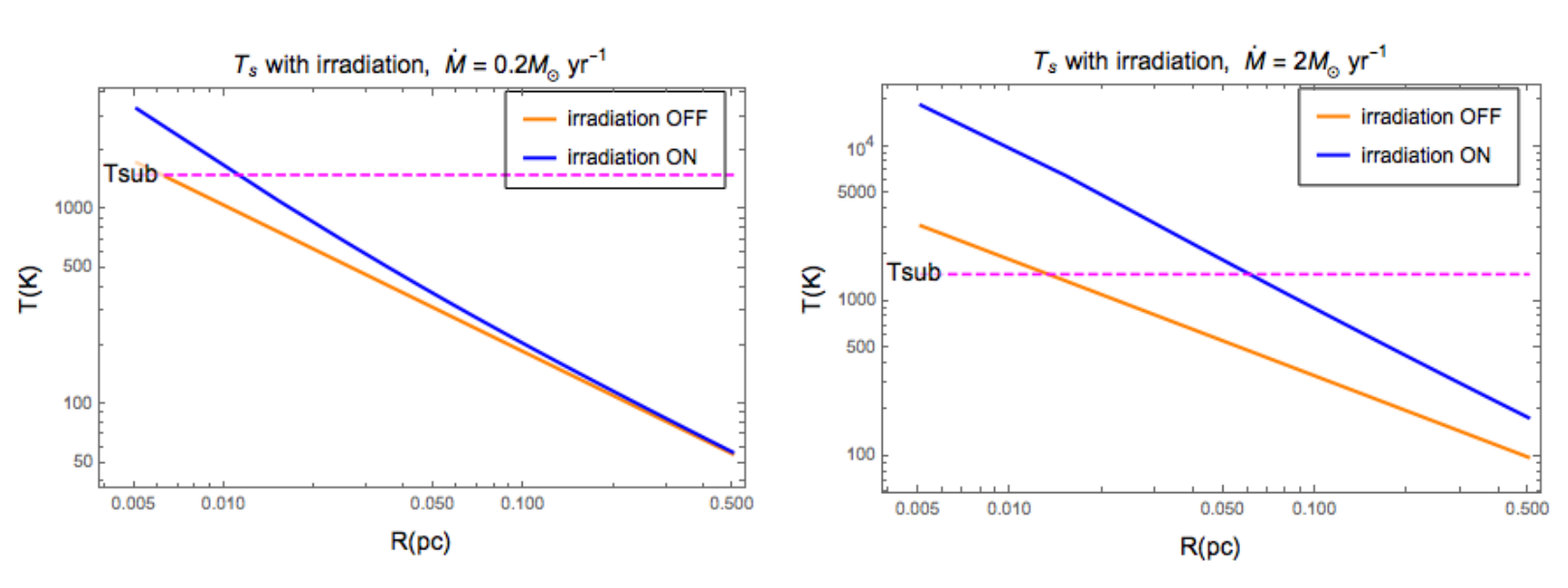}
\caption{Effect of the illumination on the surface temperature of the disk.}
\label{Fig:TcWithExtIrrad} 
\end{figure}

In presence of illumination by the radiation flux, boundary condition
for the surface temperature of the disk, $T_{s}$ should be modified:

\begin{equation}
\sigma_{{\rm B}}T_{s}^{4}=F^{+}+f(1-A)F_{n}\mbox{,}\label{eq:BCwithIrradiation}
\end{equation}

\noindent where $F_{n}$ is the component of the external flux normal to the
surface of the disk, $A$ is the disk albedo, $f$ the attenuation and angular-dependent
factor and $F^{+}$ is found from (\ref{eq:RadFlux}). Solving radiation
transfer equation (\ref{eq:dEdTau}) with (\ref{eq:BCwithIrradiation})
we have

\begin{equation}
\sigma_{{\rm B}}T^{4}=\frac{3}{4}(\tau+\frac{2}{3})F^{+}+F_{{\rm ext}}\mbox{,}\label{eq:TsolWithIllum}
\end{equation}

\noindent where $F_{{\rm ext}}=f(1-A)F_{n}$ and the photosphere is placed at
$\tau=2/3$. From (\ref{eq:TsolWithIllum}) one can see that if $F^{+}\tau_{c}\gg F_{n}$,
the mid-plane temperature, $T_{c}$ is practically independent from
external sources of heating \citep{LyutyiSunyaev1976}. The total attenuation
factor in (\ref{eq:BCwithIrradiation}) is approximately $0.5e^{-2/3}\simeq1/4$
for $A\simeq0.5$.

Vertical structure of the disk is calculated from \eqref{eq:dFdz}, and
\eqref{eq:dEdTau}, yielding vertical distribution of temperature
in the illuminated disk:

\begin{equation}
T=T_{c}\left(1-\frac{3}{2}\frac{\tau_{c}F^{+}}{T_{c}^{4}}\left(\frac{\sigma}{\sigma_{c}}\right)^{2}\right)\mbox{,}\label{eq:TcWithIllum}
\end{equation}

\noindent where we assumed the specific rate of viscous energy dissipation to
be constant, $\tau_{c}\simeq\sigma_{c}\kappa_{d}$. Surface temperature
as found from (\ref{eq:BCwithIrradiation}) can, on the other hand,
be significantly influenced by the external radiation flux:

\begin{equation}
F_{{\rm ext}}\simeq\delta(1-A)\frac{L}{4\pi r^{2}}\,\mu(2\mu+1)\mbox{,}\label{eq:FextFormula}
\end{equation}

\noindent where $L$ is calculated from (\ref{eq:AGNLum1}) and the factor $\delta\simeq H/r$
is related to the angle between the surface of the disk and the direction
of radiation flux \citep{meyerVerticalStructureAccretion1982,Spruit96}.
Estimating $\delta$ from (\ref{eq:H2R_rad}):

\begin{equation}
\delta\simeq\frac{3}{8\pi}\frac{\kappa_{d}}{cR}\dot{M}\mbox{.}\label{eq:delt_illum}
\end{equation}

\noindent The effect from the external flux is demonstrated in Figure \ref{Fig:TcWithExtIrrad}
which shows $T_{c}$ and $T_{s}$ for $A=0.5$. Not surprisingly, $T_{s}$
is noticeably influenced by irradiation.

In case of the central radiation flux, the dominating component of radiation force is directed very 
nearly along the disk surface.  The vertical radiation pressure in a single-scattering approximation, scales in the same way as the gravitational
force. Thus illumination of a thin disk by the UV flux is unlikely responsible for "puffing up" of the AGN disk at pc-scales.
X-rays penetrate much deeper, and can 
potentially lead to a much stronger 
"puffing up" due to IR pressure \citep{Chang_etal07,Dorodnitsyn16}.

The predictive power of our simple calculations of the effect of the irradiation
is limited. The attenuation of the radiation flux from the nucleus depends on
the obscuring properties of winds at smaller radii and can be addressed
only via global numerical modeling.

\subsection{Dust above the disk}\label{sec:DustAboveDisk}

Dust above the disk at $R<\rsub$ is exposed to the weathering UV and X-ray illumination from the 
nucleus.
In general, the gas-dust medium is a very efficient UV absorber.
The unattenuated UV radiation is promptly stopped in a very thin layer
where it is further converted into IR as well as heating the
gas. The UV opacity of the $0.1{\rm \mu m}$ grain is about its geometrical
cross-section, $\kappa_{{\rm UV}}\simeq10^{2}f_{{\rm d,\,0.01}}\,{\rm cm^{2}g^{-1}}$,
where $f_{{\rm d,\,0.01}}$ is gas to dust mass ration in $10^{-2}$. 
The thickness of such a ``photospheric'' layer is

\begin{equation}
\delta l_{{\rm UV}}/r_{{\rm sub}}^{{\rm UV}}\simeq5\times10^{-2}f{\rm d}_{0.01}n_{5}^{-1}L_{46}^{-1/2}\mbox{,}
\label{ConversionLayerUV}
\end{equation}
while the penetration length of
X-rays is significantly higher:

\begin{equation}
\delta l_{{\rm XR}}/\su{r}{sub}^{\rm UV}\simeq\begin{cases}
0.023, & 0.5<E<7keV\\
0.014, & E>7{\rm keV}
\end{cases}\times n_{5}^{-1}L_{46}^{-1/2},\label{ConversionLayerXRay}
\end{equation}
where $\kappa_{{\rm XR}}$ is the X-ray opacity consisting of photoionization
and Compton cross sections. In general $\delta l_{{\rm XR}}\simeq\frac{\kappa_{{\rm UV}}}{\kappa_{XR}}\delta_{{\rm UV}}.$
We adopt the photo-ionization cross-section from \citep{Maloney96}: for
$0.5<E<7$ keV we have $\sigma_{{\rm XR}}\simeq2.6\times10^{-22}\text{cm}^{2}$
and for $E>7{\rm keV}$: $\sigma_{{\rm XR}}\simeq4.4\times10^{-22}\text{cm}^{2}$.

{
Within the UV conversion layer
the UV radiation exerts pressure
of the order of $P_r\simeq0.03\, {\rm dyn\,cm^{-2}}$
on the dusty surface of the disk/torus.  
If, from within the disk, such a layer is supported 
entirely by the gas pressure, then, recalling that the equilibrium density at 
$P_r=P_g$, is  $n \simeq10^{11}{\rm cm}^{-3}$ \eqref{eq:densPgas=Prad}, 
the UV penetration length can be estimated to be just $10^{-8}-10^{-7} \su{r}{sub}^{\rm UV}$.

Illumination by
an X-ray flux has a devastating effect on the dust in an optically 
thin region. The sudden exposure of the gas-dust 
slab would create a receding evaporation layer where gas is transitioning from 
cold to $10^4-10^6$K hot component \citep[i.e.][]{Dorodnitsyn08b}.
It is not until such an X-ray flux is sufficiently
(i.e. $\tau_{{\rm xw}}\simeq1$) attenuated in the X-ray evaporative
gas/wind, when enough dust can survive, and the UV conversion layer
has a chance to actually settle.
}

The opacity of gas-dust mixture in the UV and IR is dominated by the
opacity of dust, $\kappa_{d}$, with two major contributors: silicon
at lower, and carbon at higher temperatures. Such temperature dependence
can be approximately described as

\begin{equation}\label{OpacityOfDust}
\kappa_{d}=\kappa_{0}\left(\frac{T}{T_{{\rm sub}}}\right)^{n}\;{\rm for}\;T<T_{{\rm sub}},
\end{equation}
where $\kappa_{0}=10-50\;{\rm cm}^{2}\textrm{g}^{-1}$
\citep{Semenov03}, and $n\simeq1-2$, when $T_{g}\gg T_{s}$. In general,
dust grain sublimation time-scale, $t_{{\rm sub}}$ is very short.
The mass of the dust grain can increase or decrease depending on $\Delta P=p_{{\rm vap}}-p_{i}$,
where $p_{{\rm vap}}$ is the saturation vapor pressure, and $p_{i}$
is the partial pressure of the specie, $i$ \citep{Phinney89}. The dust grain
sublimation time-scale can be estimated as

\begin{equation}
t_{{\rm sub}}=\frac{m_{{\rm gr}}}{\dot{m}_{{\rm gr}}},\label{eq:DustTimeScale}
\end{equation}where $m_{{\rm gr}}$ is the mass of a dust grain, $\dot{m}_{{\rm gr}}=4\pi a^{2}P_{{\rm vap}}\sqrt{\frac{\mu_{i}m}{2\pi kT}}$
is the dust grain mass-loss rate density, $\mu_{i}$ is the molecular
weight, and $m_{u}$ is the atomic mass unit. Since $P_{vap}\sim\exp(-{\rm few}\cdot10^{4}/T)$
it is a very sensitive function of the gas temperature. Corresponding 
time-scale is very short compared to $\su{t}{th},\,\su{t}{dyn}$.
For example,
evaluating \eqref{eq:DustTimeScale} for amorphous silicon dust, it
is just

\begin{equation}
t_{{\rm sub}}({\rm MgFeSiO_{4})}\simeq0.22\,{\rm days\mbox{,}}
\end{equation}

\noindent and $t_{{\rm sub}}\ll t_{{\rm dyn}},\su{t}{th} $ indeed follows.

\section{Discussion}\label{sec:Discussion}

It has been suggested by \citep{CzernyHryniewicz11}
that the pressure of the disk's own, local radiation on dust can drive
large-scale \textquotedbl failed\textquotedbl{} winds. \citet{BaskinLaor2018MNRAS}
recalculated the dust opacity based on the inclusion of the new data
for graphite grains. The same authors predicted in that in result
of such enhanced opacity, such a disk can \textquotedbl bulge up\textquotedbl{}
and form a compact torus 
at approximately $\text{few}\times10^{-2}$pc.
In this work we focus on a generally much broader region
where radiation pressure can impact the vertical structure of the
accretion disk in AGN. As the dusty gas spirals from galactic scales
towards the nucleus it is generally quite cold so that the disk is
very thin. Closer to the BH gas heats up due to internal  viscous
dissipation until such internally generated radiation starts to influence
the vertical structure of the disk via radiation pressure on dust grains.

The disc mid-plane temperature increases towards smaller $r$ as $T_{c}\propto r^{-9/10}\simeq r^{-1}$. 
At the radius $\Rout$ the mid-plane temperature, 
$T_{c}$ equals the temperature of dust sublimation $T_{{\rm sub}}\simeq1500$K. 
Already when $T_{c}$ reaches several hundred K, the contribution from radiation pressure 
on dust increases. When $T_{c}>\Tsub$ the mid-plane is cleared from dust and 
the total pressure at the mid-plane is dominated by that of the gas. 
However just above the mid-plane as the temperature drops to $T(z)<\Tsub$ 
opacity increases by approximately two orders of magnitude and so does the 
coupling between vertical radiation flux and the dusty gas. 
AGN accretion disks thus have two regions  where radiation pressure
is important: one close to the BH as predicted by the standard SS73
theory, and the other considerably further away, approximately at
$\Rin(T_{s}=T_{{\rm sub}})< R <\Rout(T_{c}=T_{{\rm sub}})$. 

{\mybf The mass accretion rate in ADR (ADR) does not need to
correspond to the mass accretion rate derived from the bolometric luminosity
of the nucleus, $L\propto \Mdt$ 
where $\Mdt$ is the accretion rate within inner parts of the accretion disk. The latter is approximately limited by
the Eddington accretion rate: $\dot{M}_{{\rm E}}\simeq0.2 M_{7}$
$\MsolYrM$.
The \textit{local} production rate of radiation in a disk depends on the local mass-accretion rate, 
$\mdt$. 
Correspondingly,
for the envelope of a disk to "puff up" and become slim
such disk should have the local mass-accretion rate, $\mdt > \mdt[cr]$, where 
$\mdt[cr]$ is the Eddington mass-accretion rate as calculated with respect to the opacity of dust.
Depending on assumptions about dust, $\mdt[cr] \simeq2.4-10\MsolYrM $.
If $\mdt$ in ADR is larger than $\Mdt$ the excess gas should be removed by the winds, or participate in large-scale flows. 
}


{\mybf 
In this paper we necessarily made many simplifications. 
For example, when establishing that ADR disk is highly convectively unstable we did not 
quantitatively address the problem that the
vertical structure of such a disk should be considerably altered by convection.
Instead, we assumed that in the convective layer
the convection is so efficient that it drives the equation of state to an isentropic one. In reality
we expect the total vertical flux of energy to have contributions from both convection and from radiation.
We notice though, 
that due to the very large dust opacity we expect our main conclusions to stand: ADR
is expected to be convective for  a wide range of parameters. 

{\mybf}
When considering disk irradiation and its influence on
the disk structure we only considered how such illumination changes the surface boundary condition 
for the radiation transfer problem. Such a quasi-1D approach gives qualitatively correct results but more detailed 
treatment should include angular dependent effects which can be important when disk becomes geometrically thick.

 Analogies between accretion disk physics and outflowing stellar atmospheres can be helpful.
The case of a luminous star atmosphere with high radiation pressure in continuum, in a regime 
in which the convection solution
competes with the outflowing one was calculated in \citep{BisnovatyiKogan1973StellarEnvelopesWithSupercritical}.
It was found that the increase of the radiation flux (in our case the equivalent
to the increase of $\mdt$) leads there to the
transition from convective solution to an outflowing one with some overlap between the two.  In this paper we did not calculate such outflowing
solutions, but the stellar analogy provides some evidence that such a transition may happen.  
Detailed calculations in  disk geometry are significantly more difficult, at a very minimum
requiring multi-dimensional numerical simulations. 


After the dusty gas is expelled from the disk it is exposed to radiation pressure forces from the
nucleus which can be significantly greater than the vertical radiation pressure from the disk itself. 
Depending on $\Mdt$ and $\mdt$ different types of dusty outflows can be envisaged: from thin 
layered flows along the disk surface, to the large-scale, but gravitationally
bound "failed winds", to polar hollow cone dusty outflows of different curvature. Example of such a wind is shown in Figure \ref{fig:DiskSketch}.

To produce massive outflow most efficiently two factors should align:
the local accretion rate should be greater than the local critical rate,  
$\mdt > \mdt[cr]$ and the location of ADR should be such as $\Rout\gtrsim\rsub$.
Self-regulation of accretion due to the disk's own radiation pressure on dust can 
be important for the
regulation of the SMBH growth through accretion and deserves further investigation.

}

\section{Conclusions}\label{sec:Conclusions}

 Our results can be summarized as follows:
 \begin{itemize}
\item{We have shown that there 
   is a region in an AGN accretion disk in which local radiation pressure on dust can have a major effect on the disk
    vertical structure and dynamics.}
\item{ Such an Active Dusty Region (ADR) is
approximately bounded at large radius by the dust sublimation radius
on the disk mid-plane and at small radius by the dust sublimation radius
at the disk surface.}
\item{The  outer boundary of ADR in the disk is approximately
identified as the radius, $\Rout$ where the temperature at
the disk mid-plane equals the dust sublimation temperature, $\Tsub$.
For $M_{{\rm BH}}=10^{7}M_{\odot}$ , $\Rout\simeq0.1$pc.
At $R<\Rout$ dust is cleared near the mid-plane and there
is a dramatic jump of opacity along the vertical through the disk.}
\item{ The inner boundary of ADR is located at the radius where dust
completely disappears inside the disk, i.e. at $T_{s}=\Tsub$,
where $T_{s}$ is the disk surface temperature. }
\item{ We have shown that ADR is strongly convectively unstable with significant
vertical energy transport via convection. Convection results in effective
cooling of the disk interior. It is also possible that the convection from the 
ADR provides the turbulence driver
for the BLR. }
\end{itemize}

\acknowledgements{This work was supported by NASA grant 14-ATP14-0022 through 
the Astrophysics Theory Program.}

\section*{ Appendix: Convective disk}
When energy is transported towards the surface of a
disk via convection it is often the case that $F_{{\rm conv}}\gg F_{{\rm rad}}$ where
$F_{{\rm conv}}$ and $F_{{\rm rad}}$ are convective and radiation
fluxes respectively. Convection tends to establish isentropic
distribution: $S=const.$ where $S$ is found from (\ref{eq:EntropyRad}).
Thus, to describe fully convective disk one adopts polytropic equation
of state for radiation \citet{BKBlinn77}:

\begin{equation}
P=K\rho^{4/3}\mbox{,}\label{eq:P=Kro4/3}
\end{equation}
where

\begin{equation}
K=\left(\frac{3S^{4}}{256a}\right)^{1/3}\simeq const.\label{eq:Kpolitrrad}
\end{equation}
Inserting polytropic e.s. (\ref{eq:P=Kro4/3})
equation \eqref{eq:VertBalaceTotalPNoFlux} gives:

\begin{equation}
\frac{dP}{dz}=-\Omega^{2}K^{-3/4}zP^{3/4}\mbox{.}\label{eq:dPdZAdiab}
\end{equation}
Solving further this equation, one can eventually obtain the following
simple relations:

\begin{eqnarray}
\rho & \simeq & \rho_c\left(1-\frac{z^{2}}{z_{b}}\right)^{3}\mbox{,}\label{eq:roSol}\\
P & \simeq & P_c\left(1-\frac{z^{2}}{z_{b}}\right)^{4}\mbox{,}\label{eq:PresSol}
\end{eqnarray}

When simplifying (\ref{eq:roSol}),(\ref{eq:PresSol}) we took into account that $P_{b}\ll P_{c}$,
$\rho_{b}\ll\rho_{c}$ and adopting {\it these relations} for simplicity the assumptions: $P_{b}\simeq0$, $\rho_{b}\simeq 0$
at $z=z_{b}$, where $\rho_{b}$ and $P_{b}$ are the corresponding values at the boundary of the disk.
A surface boundary condition follows from  integrating \eqref{eq:dPdz} between $z_b$ and infinity:
$P_b=z_b\Omega^2\tau_b/\kappa_d$, where $\tau_b\simeq 2/3$, and all properties of a polytropic disk can then
be derived as in \citep{BKBlinn77}.

\newpage
\section{Glossary}

\begin{longtable}{c l}  
Symbol & description, Sec., (eq. number) \\
\hline
\text{ADR} & "Active Dusty Region", \ref{sec:Introduction}\\
L & total luminosity, Sec.\ref{sec:GlobalParameters}, \eqref{eq:AGNLum1}\\
$M$  &  mass of the BH, Sec.\ref{sec:GlobalParameters} \\
$\epsilon$ & accretion efficiency, Sec.\ref{sec:GlobalParameters}, \eqref{eq:AGNLum1}\\
$\Mdt$ & mass-accretion rate near BH, Sec.\ref{sec:GlobalParameters}, \eqref{eq:AGNLum1}\\
$L_{{\rm E}}$ & Eddington luminosity, Sec.\ref{sec:GlobalParameters}, \eqref{eq:EddLum1}\\
$\kappa_{e}$ & electron opacity, ${\rm cm^{2}g^{-1}}$, Sec.\ref{sec:GlobalParameters} \\
$R$ & radius in physical units, Sec.\ref{sec:GlobalParameters}\\
$r$ & radius in scaled units, Sec.\ref{sec:GlobalParameters}\\
$\su{R}{AGN}$ & "outer radius of AGN", Sec.\ref{sec:GlobalParameters}, \eqref{eq:rAGN} \\
$\su{\sigma}{Blg}$ & bulge stellar velocity dispersion, Sec.\ref{sec:GlobalParameters}, \eqref{eq:rAGN} \\
$R_g$ & Schwarzschild radius, Sec.\ref{sec:GlobalParameters}, \eqref{eq:RadSchwarz}  \\
$\Omega$ & angular velocity in the disk \\
$T$ &  gas temperature, Sec.\ref{sec:GlobalParameters} \\
$\rho $ &  gas density, Sec.\ref{sec:GlobalParameters} \\
$n $ &  gas number density, Sec.\ref{sec:GlobalParameters} \\
$H$ & thickness of AGN disk, Sec.\ref{sec:GlobalParameters}, \eqref{eq:H2R_at_rAGN} \\
$F_{{\rm ext}}$ & flux from the nucleus, \ref{sec:GlobalParameters}, \eqref{eq:FluxAGNanisotr} \\
$F_{{\rm loc}}$ & disk local radiation flux, Sec.\ref{sec:GlobalParameters}, \eqref{eq:DiskLocFlux} \\
$F_{{\rm tot}}$ & total vertical energy flux in a disk, \ref{sec:Discussion}\\
$\theta$ & inclination angle from the normal to the disk, Sec.\ref{sec:GlobalParameters} \\
$\mu$ & $\cos\theta$, Sec.\ref{sec:GlobalParameters}, \eqref{eq:FluxAGNanisotr} \\
$f(\theta)$ & angular dependence of the radiation flux, Sec.\ref{sec:GlobalParameters}, \eqref{eq:FluxAGNanisotr} \\
$\mdt$ & local mass-accretion rate near BH, Sec.\ref{sec:GlobalParameters}, \eqref{eq:DiskLocFlux} \\
${\bf g}_\txt{rad}$ & radiation pressure vector, Sec.\ref{sec:GlobalParameters} \\
$\su{n}{eq}$ &  density at which $P_g=P_r$, Sec.\ref{sec:GlobalParameters}, \eqref{eq:densPgas=Prad}\\
$P_g$ & gas pressure, Sec.\ref{sec:GlobalParameters}, \eqref{eq:Pgas} \\
$\su{{\cal R}}{gas}$ & gas constant, Sec.\ref{sec:GlobalParameters} \\
$\mu_m$ & mean molecular weight, Sec.\ref{sec:GlobalParameters}, \eqref{eq:Pgas} \\
${\cal R}$ & ${\cal R}/\mu_m$ -modified gas constant, Sec.\ref{sec:GlobalParameters} \\
$P_r$ & radiation pressure, Sec.\ref{sec:GlobalParameters}, \eqref{eq:Prad} \\
$a$ & radiation constant, Sec.\ref{sec:GlobalParameters}, \eqref{eq:Prad} \\
$t_{{\rm dyn}}$ & free-fall time-scale, Sec.\ref{sec:GlobalParameters}\\
$v_r$ & radial velocity, Sec.\ref{sec:GlobalParameters}\\
$\alpha$ & viscosity parameter, Sec.\ref{sec:GlobalParameters}\\
$t_{{\rm a}}$ & disk accretion time-scale, Sec.\ref{sec:GlobalParameters}, \eqref{eq:t_visc}\\
$t_{{\rm visc}}$ & disk viscous time-scale, Sec.\ref{sec:GlobalParameters}, \eqref{eq:t_visc}\\
$t_{{\rm th}}$ & disk thermal time-scale, Sec.\ref{sec:GlobalParameters}\\
$\kappa$ & opacity of the accreting material, Sec.\ref{sec:DiskThicknessWithRadPressure}, \eqref{eq:dPdz}\\
$\kappa_d$ & dust opacity, Sec.\ref{sec:DiskThicknessWithRadPressure}, \eqref{OpacityOfDust}\\
$\su{\kappa}{UV}$ & UV dust opacity, Sec.\ref{sec:DustAboveDisk}\\
$\su{\kappa}{XR}$ & X-ray gas opacity, Sec.\ref{sec:DustAboveDisk}\\
$\su{\sigma}{XR}$ & X-ray gas cross-section opacity, Sec.\ref{sec:DustAboveDisk}\\
$g_{z}$ & vertical gravitational acceleration, Sec.\ref{sec:DiskThicknessWithRadPressure}, \eqref{eq:gz}\\
$\Omega_{K}$ & Keplerian angular velocity, Sec.\ref{sec:DiskThicknessWithRadPressure}, \eqref{eq:OmegaKepler}\\
$\mdtcr$ & Eddington accretion rate for dust opacity, Sec.\ref{sec:DiskThicknessWithRadPressure}, \eqref{eq:MdotCritDust}\\
$T_{{\rm vir,r}}$ & ``virial'' temperature for the radiation dominated medium, Sec.\ref{sec:DiskThicknessWithRadPressure}, \eqref{eq:TvirRad}\\
$T_s$ & temperature at the disk surface, Sec.\ref{sec:DustSubRegIndisk}, \eqref{eq:Teff}\\
$T_c$ & mid-plane temperature, Sec.\ref{sec:DustSubRegIndisk}, \eqref{eq:TcAnalytPgas}\\
$\Sigma_c$ & disk surface density, Sec.\ref{sec:DustSubRegIndisk}\\
$\tau_c$ & mid-plane optical depth of the disk, \ref{sec:DustSubRegIndisk}\\
$\tau_{\rm phot}$ & optical depth at the disk photosphere, Sec.\ref{sec:DustSubRegIndisk}\\
$\su{\sigma}{B}$ & Stefan-Boltzmann constant, Sec.\ref{sec:DustSubRegIndisk}\\
$\su{R}{in}$ & inner dust sublimation radius in the disk Sec.\ref{sec:DustSubRegIndisk}, \eqref{eq:Rin}\\
$\su{R}{out}$ & outer dust sublimation radius in the disk Sec.\ref{sec:DustSubRegIndisk}, \eqref{eq:Rout}\\
$\su{R}{sub}$ & global dust sublimation radius Sec.\ref{sec:DustSubRegIndisk}, \eqref{eq:RsubAGN} \\
$P$ & total pressure Sec.\ref{sec:Solution}, \eqref{eq:Ptot} \\
$\Sigma$ & surface density, Sec.\ref{sec:Solution}, \eqref{eq:SigmaS}\\
$F$ & vertical radiation flux, Sec.\ref{sec:Solution}, \eqref{eq:dFdz}\\
$F^+$ & vertical radiation flux from the surface , Sec.\ref{sec:Solution}, \eqref{eq:dFdz}\\
$F_{{\rm conv}}$ & convective flux Sec.\ref{sec:Convection}, \eqref{eq:Fconv}\\
$\nu$ &  effective viscosity , Sec.\ref{sec:Solution}, \eqref{eq:Nu_HTSigma}\\
$q_{{\rm v}}$ & specific rate of viscous energy dissipation, Sec.\ref{sec:Solution}, \eqref{eq:qv(specific)}\\
$\sigma$ & mass coordinate , Sec.\ref{sec:Solution} \eqref{eq:MassCoordinate}\\
$H_g$ & scale-height of the equatorial gas layer of the disk, Sec.\ref{sec:SolWithRadPressure}, \eqref{Hgas}\\
$\su{T}{c,gas}$ $\Tcgas$ \\
$\left(\frac{dT}{dz}\right)_{{\rm ad}}$ & adiabatic temperature gradient,Sec.\ref{sec:Convection}, \eqref{eq:dTdz_ad_num}\\
$\left(\frac{dT}{dz}\right)_{{\rm rad}}$ & radiative temperature gradient,Sec.\ref{sec:Convection}, \eqref{eq:dTdz_ad_num}\\
$S_{r}$ & entropy of the radiation gas Sec.\ref{sec:Convection}, \eqref{eq:EntropyRad}\\
$C_{p}$ & heat capacity at constant pressure, Sec.\ref{sec:Convection}, \eqref{eq:Cp}\\
$\Delta\nabla T$  & temperature excess of the convective element, Sec.\ref{sec:Convection}, \eqref{eq:DelNablaT}\\
$l$  & mixing length, Sec.\ref{sec:Convection}\\
$\epsilon_{0}$ & mixing length parameter, Sec.\ref{sec:Convection}\\
$F_n$ &component of the external flux normal to the 
surface of the disk, Sec.\ref{sec:ExternalIrradiation}, \eqref{eq:BCwithIrradiation}\\
$A$ & disk albedo, Sec.\ref{sec:ExternalIrradiation}, \eqref{eq:BCwithIrradiation}\\
$\delta$ & angle between the surface of the disk and the direction
of radiation flux, Sec.\ref{sec:ExternalIrradiation}, \eqref{eq:FextFormula}\\        
$\delta l_{{\rm UV}}$ & thickness of the UV conversion layer, Sec.\ref{sec:DustAboveDisk},
\eqref{ConversionLayerUV}\\
$\delta l_{{\rm XR}}$ & thickness of the X-ray conversion layer, Sec.\ref{sec:DustAboveDisk},
\eqref{ConversionLayerXRay}\\
$t_{{\rm sub}}$ & dust grain sublimation time-scale, Sec.\ref{sec:DustAboveDisk}, \eqref{eq:DustTimeScale}\\
\end{longtable}


\bibliography{Dorod}

\end{document}